# Edge Computing for Computer Vision in IoT: Feasibility and Directions

Panagiotis Savvidis[1*], George A. Papakostas[1]

[1]Department of Computer Science, Democritus University of Thrace, Kavala 65404, Greece

## Abstract

The convergence of decentralized architectures integrating Machine Learning, Computer Vision and Low Power Wide Area Networks is increasingly becoming an integral part of our daily existence. Internet of Things serves as a real-time data conduit enhancing decision making via embedded technology and continuous data exchange. This paper explores the feasibility of Edge Computing as a foundational pillar in this evolving landscape. We experiment under real world, dynamic conditions, evaluate the technological aspects, strategies, process flows and key observations under the broad Edge Computing domain. Research pathways include Multi-access Edge topologies in future 6G networks, model quantization, and satellite-enhanced communication platforms. Additionally, a discussion is added supporting the advanced AI functionalities, including zero-shot learning, multi modal perception, and decentralized generative AI, thereby broadening the scope of intelligent applications across various domains. The significance and research objective of this study are threefold: (1) evaluation of LoRaWAN and satellite IoT communication strategies, (2) analysis of CV workloads on edge hardware and (3) future research directions where Edge Computing can support low-latency, energy-efficient and socially impactful IoT applications. By explicitly addressing these aspects, we aim to establish a clear link between the technological feasibility, ultimately with a practical and socioeconomic relevance.

*Corresponding author. Email address: savidis.panagiotis@gmail.com

## 1. Introduction

The Internet of Things (IoT) is a global network of physical nodes that collect and share data. Connecting all these different objects with the added built-in sensors, enables a real time communication without involving the human intervention. Consequently, Edge Computing (EC) is the focus area where centralized topologies such as cloud and central servers are reduced from the data processing, analysis and computation burdens. The raw data produced by the numerous things can be enormous that could infer cloud approaches and conventional computing less efficient to handle [1],[2]. Researchers in [3] are motivated by the Systematic Literature Review (SLR) to analyze resource estimation, consolidation, load balancing and computational offloading methods that define Edge and Fog topologies. Thereafter, the scope is to address the massive IoT data by being consumed at the network edge. Embedded boards are encapsulating Machine Learning (ML) and IoT capabilities, smaller in dimensions and more compute capable. They utilized in a variety of purposes such as system development, robotics, education and others. The context of Artificial Intelligence (AI) and Deep Learning (DL), led the Single Board Computer (SBC) manufacturers to compete on small dimensions and low power draw offerings. In telemetry terms and the exception of Internet Service Providers (ISP), Low Power Wide Area Networks (LPWAN) adaptations such as Nb-IoT, Sigfox or LoRaWAN are the obvious evolution. This paper would provide the empirical evidence through: (i) LoRaWAN Field testing in urban, suburban, and rural environments and practice with satellite IoT via TinyGS, (ii) Benchmarking YOLOv4 object detection on the Raspberry Pi 4 and Jetson Nano, measuring inference time, system load, and energy consumption (iii) demonstrating the future directions for multimodal Edge AI, model quantization and highlighting the socioeconomic dynamics of Edge based CV deployments. To our knowledge, we demonstrate the first integrated study combining CV workloads on low power SBCs and LPWAN into a unified feasibility analysis. By clarifying these objectives, this study positions itself at the IoT, AI and next generation networking intersection, contributing on both the academic discourse and practical designs of future EC systems. Structuring this work, we unfold in section 2 the baseline characteristics for EC, in section 3 introduction on LPWANs, LoRaWAN metrics, SatIoT specifics and a practical attempt with TinyGS as an extension from [4]. Section 4 is focused on Computer Vision (CV) literature review, image annotation, datasets, cameras, the performance evaluation of two SBCs and YoloV4 classifier performance metrics. Section 5 is dedicated on the discussion, implementation insights, feasibility considerations and key contributions.

## 2. Edge Computing Formal Observations and case studies

A practical example for a decentralized topology is a flying plane. Among its sensors and other systems an estimated 300 GB generation of data is a typical occurrence [5]. Continuous ground communication or direct satellite data link is less feasible for economic and practical reasons [6], but attainable under certain strategies [7]. Vast data can be processed and stored onboard with aggregated telemetry towards a ground station. Over the Top (OTT) platforms can bottleneck centralized topologies, hence the development of Multi Access Edge topologies (MEC). These are the cases of service providers moving the workloads and services towards the network Edge and out of the core establishments or data centers [8]. A contribution by [9] on intelligent transportation systems pertain the feasibility aspects for Edge AI in IoT. The local and within the field data process will favor latency reduction, positive impact on the Quality of services (QoS) and simplification. In [10] computation offloading is highlighted where Edge-IoT nodes will connect the physical and digital environments that would empower businesses, add productivity and help for informed decisions and actions. Authors at [11] portraying the utilization of demanding Augmented Reality (AR) applications with EC. The requirement is continuous object detection in a wide Field of View (FoV) with high frame rates and latency minimization. EC will in fact strengthen the overall experience by local computations while avoiding unnecessary transmissions to the cloud. Authors at [12] are evaluating strategies to reduce wireless bandwidth demands for a drone video exchange and live streams. Another work in [13] gives the insights of Cloud Computing advantages but highlight that the on-premises computing is expected to save on management costs if implemented under an organized Total Cost of Ownership (TCO) approach. Additionally, in [14] the authors compare the management and processing costs between a Cloud and hybrid Edge-Cloud approach, monitoring sensors and a camera feed of a wind farm over the distance of 200 miles. This comparison showed a 95% data traffic optimization favoring the hybrid Edge-Cloud proposition with a cost reduction on the triennial study from 81000$ to 29000$, a ⅔ decrease. Similar in [15] and the contribution for smart home monitoring, the popular Raspberry SBC is utilized for data analysis and actuation commands by keeping communication and computation costs low. Another paper in [16] promotes the maritime EC for ocean digitization, intelligence and communication expansion cases. Authors verifying the feasibility of the proposition by the integration of a Jetson Xavier NX, a camera and sensors to process and objectify the recognized object. To follow up we summarize the benefits of EC:

**Latency Reduction**: where local ML inference enables fast, responsive decisions. [17].
**Cost Reduction**: less raw data transmission minimizing communication expenses. [10].
**Privacy & Security**: decentralized data handling reduces leakage risks. [18],[6].
**Independent operation**: devices can store/manage decisions even offline.
**Sustainability**: same hardware reused for multiple applications. [19].
**Agility**: rapid prototyping with low-cost, flexible hardware.

### 2.1 Edge Computing architecture overview

In general consensus inexpensive SBCs with low power requirements are introducing EC capabilities regardless of their limited computational potential. In telemetry terms, IoT is adapted with Nb-IoT, Sigfox and LoRaWAN, Fig 1 depicts the various layers of the concept.

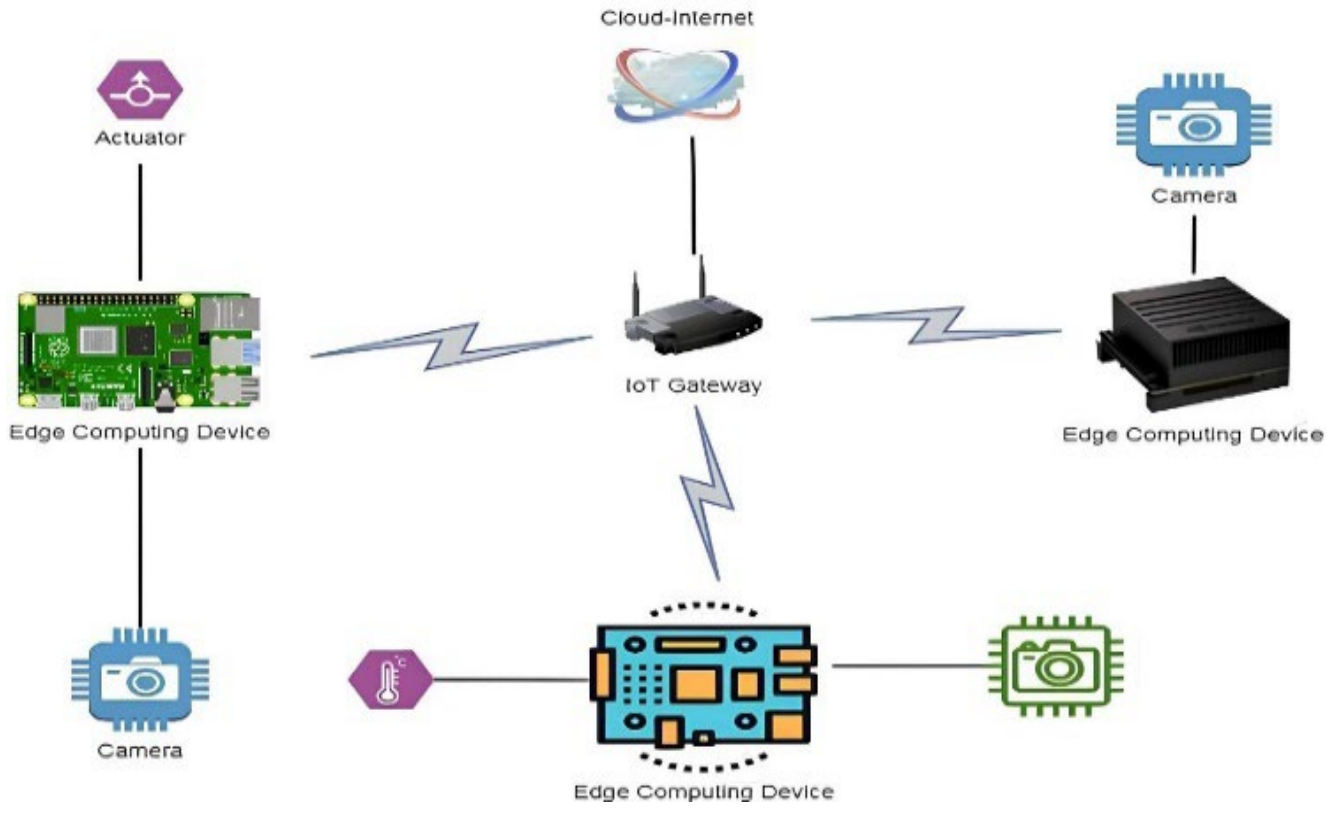

**Figure 1.** Embedded devices contained to the Edge of the network, communicating with the internet.

Also, GPU accelerated examples in bibliography are ensembling this architecture. Traditionally, sensors and actuators at their core foundation are computationally incapable, thus they will feed a central decision-making entity as being the heart of the system. This middle layer can consist of a field programmable gate array (FPGA), an SBC or an Application Specific Integrated Circuit (ASIC). Additionally, utilization of Neural Compute Stick (NCS), Vision Processing Unit (VPU) and Graphic Processing Unit (GPU) are accelerating neural computations and real time AI inference capabilities [20]. Another component of these platforms is the communication layer of either the LPWAN or mobile data. The benefit of processing raw data made by various sensors or cameras in a closed loop offers the chance of reduced yet significant metadata to be transmitted. Author at [6] paraphrases the concept as: "***In data abstraction those raw elements are consumed within the device, hence the human involvement to the data is minimized but proactive***".

### 2.1.1 Edge Computing use cases

**Computation Offloading (CO):** Will be the process of linking computer intensive tasks and storage to a different co-processor or to an Edge topology. These architectures perform communication and computational resources allocation or load balancing [10]. In [21] they propose Deep Neural Networks (DNN) offloading strategies to the Edge on supporting smart IoT.
**Smart Home:** It is the derivative for smart home applications with seamless communications and contained operations focusing on privacy and security. A home with various connected devices is producing a fair amount of data, with this paradigm. EdgeOS has been introduced by utilizing Edge Routers [22].
**Smart City:** To a much larger extent applications already serve urban, suburban and rural areas producing intimidating volumes of big data. As IoT integrates infrastructures such as smart utility meters, Intelligent Transportation Systems (ITS), healthcare, public safety, and farming, Edge topology can mitigate these loads from centralized infrastructures.
**Collaborative Edge Computing (CEC):** This is an ad hoc style communication to facilitate collaboration of multiple EC hosts for the purpose of data sharing across geographically distributed heterogeneous devices [23]. Referred to as early information exchange systems between nearby Edge stations. This technique can resemble the Cloud operation on a smaller scale, thus inheriting some of its drawbacks through the cost of data transmission.
**Industrial Analytics:** Data leverage from sensors, machines and other connected devices will improve efficiency, and predict potential issues before they occur. By applying advanced analytics techniques such as ML and AI to industrial data, organizations can gain valuable insights that drive decision-making. Overall, edge analytics can help businesses save money, while also improving the quality of their analytical models [24].

**Edge AI:** By combining EC and AI, we reach another subject and purpose. They bring analysis, computation and decision making closer to the data sources. Silicon on Chip (SoC) manufacturers are developing newer designs, enabling system engineers to perform AI and ML tasks in the network edge.
Beyond these generalized cases, a real-world deployment by [25], showcases the significance of EC as it aims to protect farming crops. The prototype is designed to function as an intelligent animal repelling system that recognize wild nature species. The system's functionality is enabled by Raspberry, or a Jetson Nano and Yolo CV algorithm. For serving the rural communication purposes LoRaWAN and Xbee radio were the most suitable solutions between the SBC and the sensors.

Table 1. Edge and Cloud characteristics

| | Edge Computing | Cloud Computing |
|---|---|---|
| Architecture | Decentralized-Local | Centralized |
| Data Processing | Directly from the source | Away of the source |
| Latency | Minimized | High |
| Connectivity Requirements | Various protocols-LPWAN | High speed internet |
| Computing capabilities | Low | High |
| Naturalization-Infrastructure | Growing | High |
| Analysis | Short term | Long term |
| Cost/Data throughput | Lower | Higher |
| Energy efficiency | Better | Inferior |
| Privacy-Prone to cyber attack | Better | Inferior |

### 2.1.2 Challenges in Edge Computing

Due to the nature of heterogeneous computing platforms, maintenance, revision and troubleshooting in the remoteness of the nodes becomes a complex task, the need of unique device identifiers for logistical purposes is crucial for that matter. Standardization of different communication interfaces and devices is a prerequisite. The lack of technical proficient personnel to release changes and overcome the technical barriers or complexities is also an issue. Based on observations gathered under the broad EC term, a work by [26] outlines the performance metrics on Edge and Cloud. As previously discussed, each of the specified approaches are based on certain application criteria. An argument arises about the inversion performance issue for Edge Computing. It describes the high queuing delays endured in Edge topology caused by the less potent hardware on the end-to-end comparison with the cloud for moderate loads. Performance metrics between the two paradigms had been compared using queuing models in order to bring a common ground on the latencies for different workloads. While a moderate portion of EC implementations might not utilize broadband communications or having on par hardware capabilities, a direct comparison might not be justified. The advantages of Edge architecture are not reflected due to generalization of dispersed Cloudlets and Micro datacenter applications. With that notion a cloudlet defines computation offloading, where in fact Edge is the basis of keeping the load decentralized. The reported mean and tail latency for Edge applications in moderate loads should be scaled proportionally for both architectures. Another study in [27] proposing the methodologies towards the sustainability of medium sized EC deployments and a use case for Advanced Driver Assistance Systems (ADAS). Indicative is the Power Usage Effectiveness (PUE) for EC, deviating from the cloud farm index respectively. Energy efficiency and collaborative orchestration are the strategies to leverage any implications that dense EC collocated solutions are inherent to. By summarizing, Edge and Cloud solutions aren’t challenging each other, but deployable propositions for organizations and developers to identify the needs, cost requirements and assessing what works the best.

## 3. LPWAN options

Researching the IoT communication spectrum between the end nodes, we dedicating the following comparison (Table 2) on different LPWANs and LoRaWAN [28]. Sigfox excels in energy efficiency but with limited capacity, NB-IoT benefits from LTE infrastructure at higher power costs [29], and IEEE 802.15.4 offers reliability but with a short range [30]. In our opinion, LoRaWAN strikes the balance between long range coverage, low energy draw, moderate data rates and easy integration. From the developer standpoint, multitude of resources are available for experimenting. These make LoRaWAN a popular choice for various IoT deployments.

Table 2. Characteristics of common LPWAN options

| Sigfox | LoRaWAN | NB-IoT | IEEE 802.15.4 |
|---|---|---|---|
| 1 Un-licensed ISM bands-868, 915 MHz and 433 MHz | Un-licensed ISM bands-868, 915 MHz and 433 MHz | Licensed LTE bands | Un-licensed ISM bands-868 MHz, 915 MHz (low band) and 2.4 GHz (high band) |
| 2 D-BPSK | CSS | QPSK | BPSK / O-QPSK |
| 3 100 Hz | 250 kHz and 125 kHz | 200 kHz | 2 MHz high band |
| 4 100 bps | 50 kbps | 200 kbps | 250kbps (2.4GHz) |
| 5 12 bytes uplink, 8 bytes downlink | 222 bytes | 1600 bytes | 127 bytes |
| 6 140 uplinks, 4 downlink | 30 seconds of uplink per device | Unlimited | Unlimited |
| 7 10 km (urban), 40 km (rural) | 5 km (urban), 20 km (rural) | 1 km (urban), 10 km (rural) | 10-75m, 1000m in LOS |
| 8 High | Low | Low | High |
| 9 High | High on Class A-Low in Class C | Low | High |
| 10 No | Yes | No | No |
| 11 Low | Low | Moderate | High |
| 12 No | AES 128b | LTE encryption | AES 128b |
| 13 No | Yes | No | No |
| 14 ETSI | LoRa-Alliance | 3GPP | IEEE |

*1.Physical layer (PHY), 2. Modulation, 3. Bandwidth, 4. Maximum data rate, 5. Maximum payload length, 6. Maximum messages-daily, 7. Range, 8. Interference, 9. Latency, 10. Adaptive data rate, 11. Energy Drawn, 12. Authentication, 13. Private network, 14. Standardization.*

## 3.1 Background on LoRaWAN

Is a networking protocol enabling long-range (in rural and LOS) transmissions of more than 15km. The gateways relaying messages between the end nodes and network servers in RF traffic and IP packets respectively targeting wireless battery-operated things Fig 2. The things or nodes are referred to literature as Wireless Sensor Networks (WSN). They listen and forward broadcasts in a license free sub-Gigahertz RF with bands like 433, 868, 915, 470 and 923 MHz's.

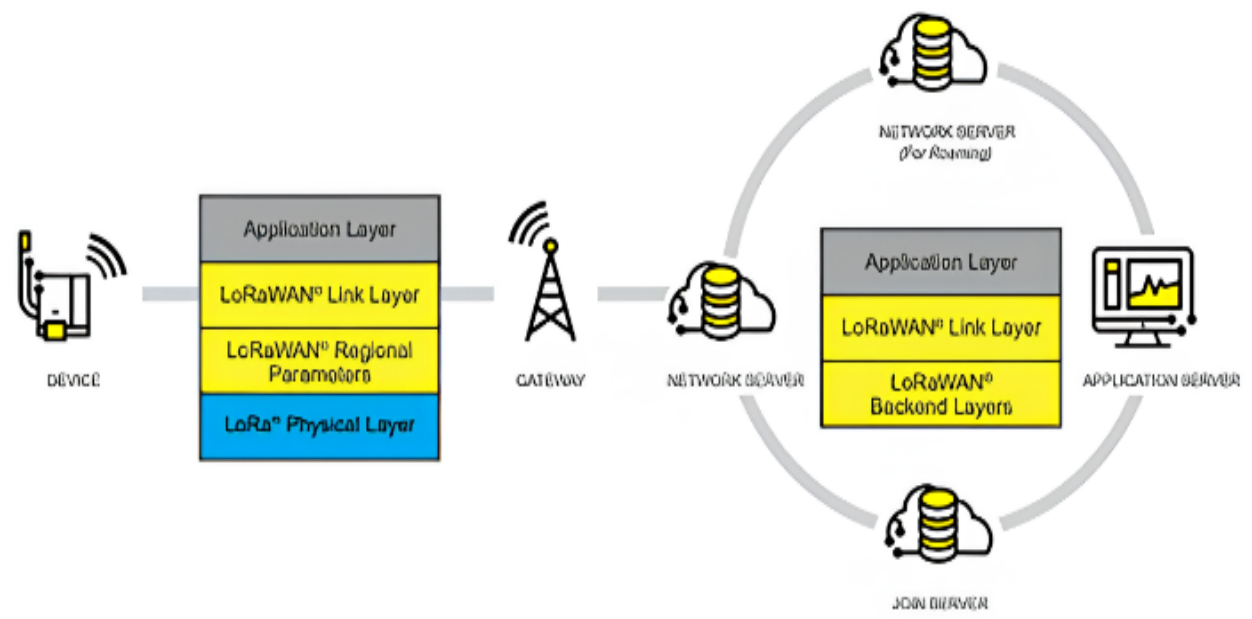

**Figure 2.** LoRaWAN Architecture

Lora is a modulation method based on Chirp Spread Spectrum (CSS). It stands for Compressed High Intensity Radar Pulse, a signal whose frequency increases or decreases over time. Advantages of CSS radio modulation is interference resilience, a good link budget and low power characteristics. The physical layer of LoRa consists of preamble up chirps, frame delimiter down chirps and the varying length up chirps that represent the data, Fig 3.

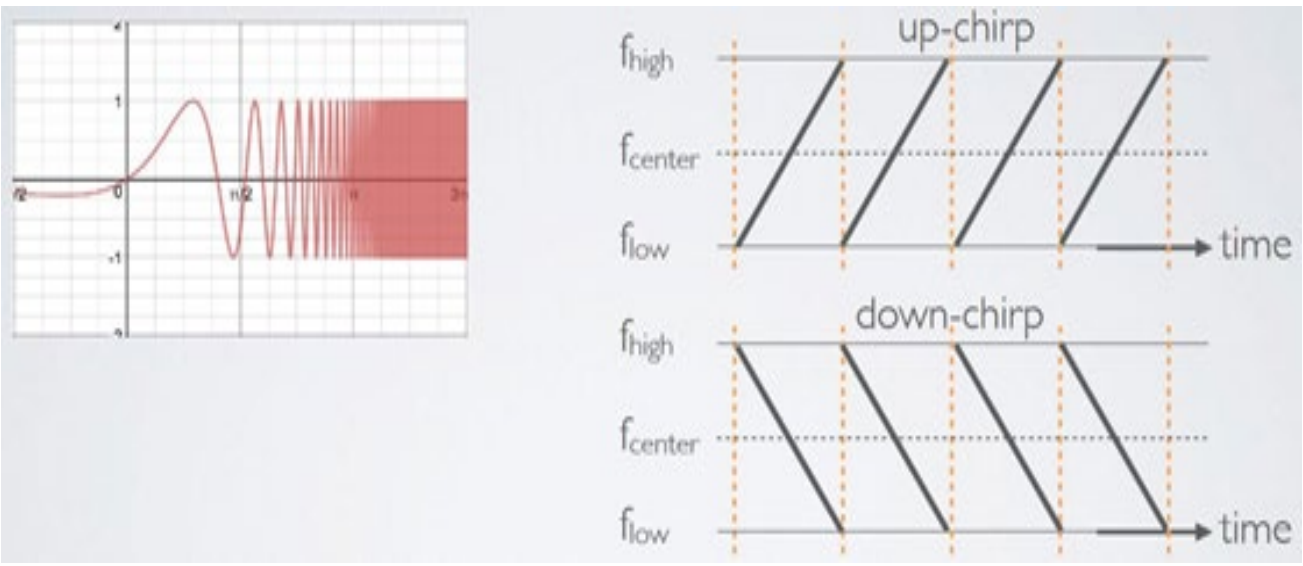

**Figure 3.** The chirp spread spectrum modulation technique

**A**, **B** or **C** are class definitions for the downlink receive windows intervals, an **A** device class strives for efficiency by receiving the downlink after an uplink before a sleep state. On the contrary the lowest latency is reserved to class **C** devices and an open receiver state, a mode for non-battery-operated nodes.

**Adaptive Data Rate** (ADR) is the mechanism that assesses the available SNR margin to increase the data rate. The mechanism adjusts the spreading factor, transmission power and bandwidth values that benefits the airtime and energy consumption.

**Spreading Factor** (SF) relates to the number of chips to represent a symbol with an exponential factor of $2^{SF} = 1$ symbol. More chips lead to a wider distance signal reach and gain but with longer airtime. SF7 holds 128 where SF12 4096 chirps per symbol.

**Code Rate** (CR) on a low SF setting will retain additional redundancy with a Forward Error Correction (**FEC**), a strategy towards link interferences. Other adjustable Data Rate indices are the bandwidth (BW) in kHz and transmission (TX) power in dBm.

Table 3. SF to airtime comparison based on 11 bytes plus the overhead payload

| SF7B W125 | SF8B W125 | SF9B W125 | SF10B W125 | SF11B W125 | SF12B W125 |
|---|---|---|---|---|---|
| 61.7 ms | 113.2 ms | 205.8 ms | 370.7 ms | 823.3 ms | 1.482 ms |

### 3.1.1 LPWAN metrics methodology

With this assessment urban, suburban and rural environments behavior are examined. For the tests four gateways are utilized, with C and D being within the urban environment while in the city outskirts the gateways B and A, in Fig 4.

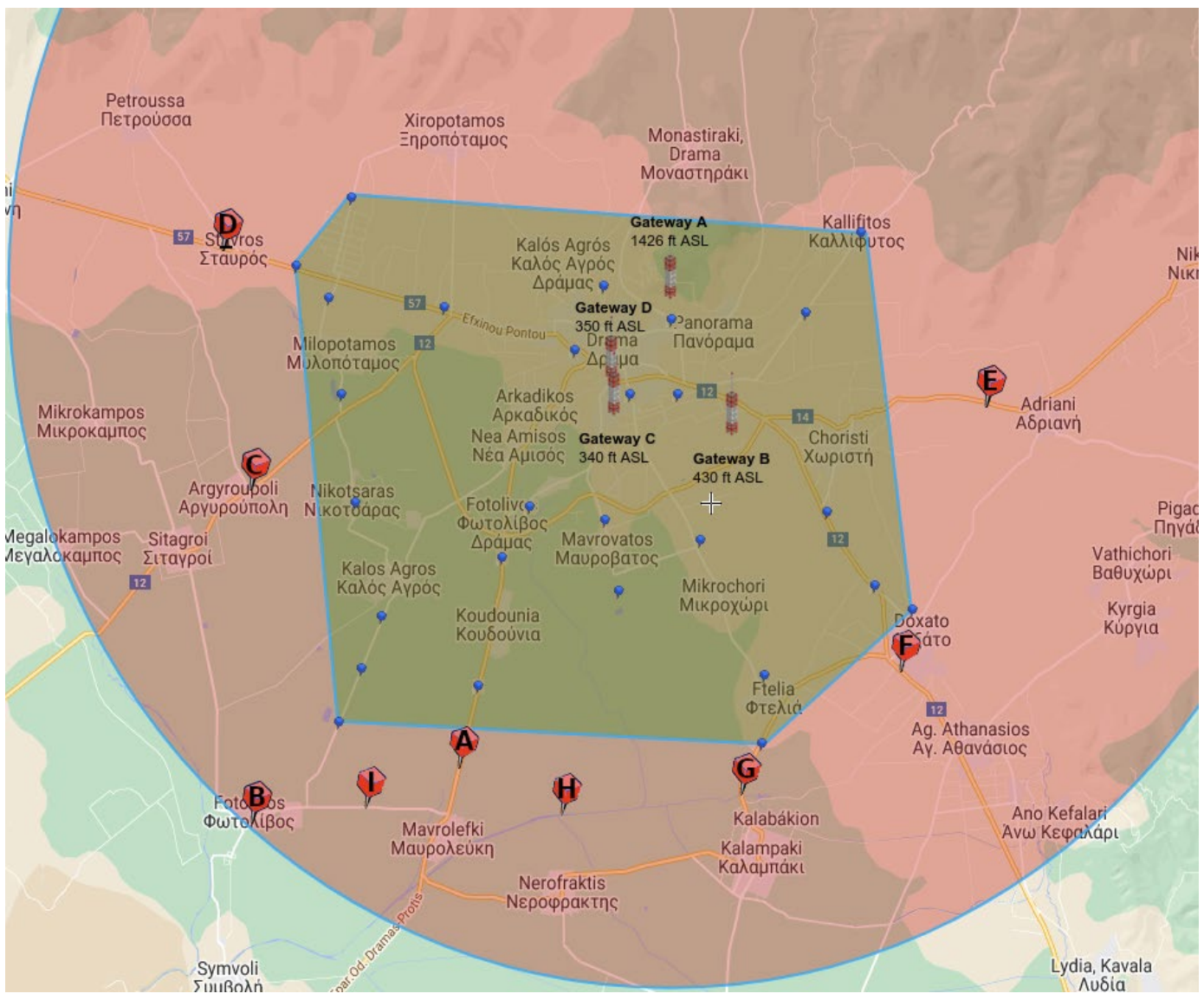

**Figure 4.** The 4 gateways along with the 27 test spots (blue pushpins) and 9 rural test spots (red push pins)

An academic contribution in [31] concludes that LoRaWAN reliable communications are based on strategic placement of the Ground Stations (GS). A key feature to base conclusions is the SF potential in different test spots.

With SF we can directly determine the network coverage, blind spots and indicate energy demands as a derivative of payload airtime. Inside this airtime envelope an optimized duty cycle of the radio links is regulated.

### 3.1.2 Urban and suburban communications

The green overlaid polygon from Fig 4 covers an area of 143.75 $km^2$ and is defined as the radio link quality evaluation for urban and suburban areas. The blue pins are the test spots and assessed on the Received Signal Strength Indicator (RSSI) and Signal to Noise Ratio (SNR) at various SF modes. For urban and suburban environments RSSI and SNR were at their best levels at (**-87dBm and 12,5 dBm**) with their worst at (**-126dBm and -14dBm**) respectively. With all the SF values from 27 test spots a mean performance of (**-110dBm and 1.44dBm**) for RSSI and SNR was obtained. Worth mentioning that multicast uplinks are possible, where the same payload can be listened to by 3 stations with values shown on Table 4.

Table 4. A multicast reception example

| Gateway ID | RSSI@SF10 | SNR@SF10 |
|---|---|---|
| D / B / A | -126 / -112 / -90 | -14 / -12.5 / 8.5 |

Out of the 3 stations D received a weak uplink and A had the best reception.

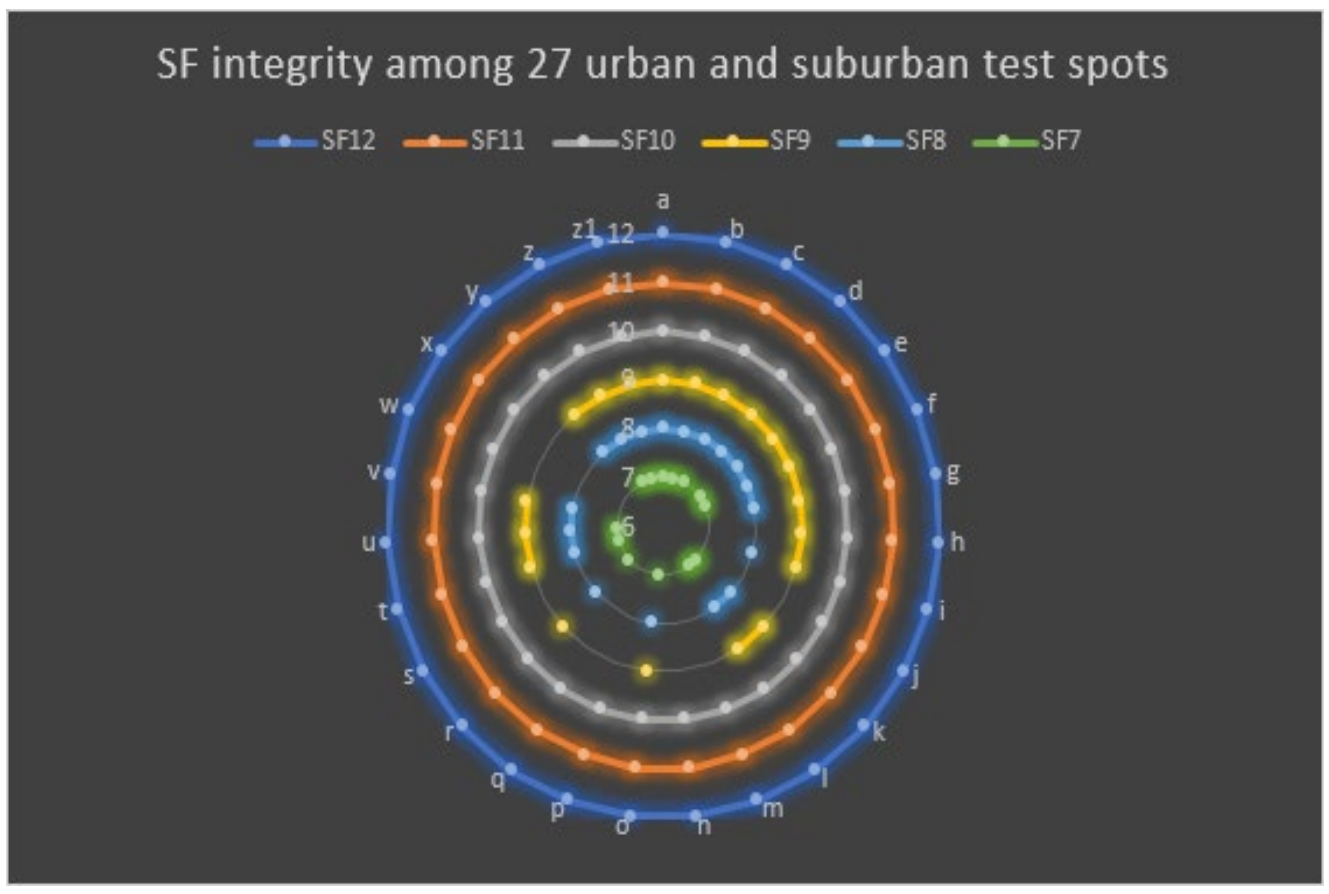

**Figure 5.** Prevalence of SF modes on urban and suburban environments.

Fig 5 above shows the complete coverage on SF modes 12, 11 and 10, attributed from the high radio sensitivity and the longer chirp durations. In modes 9, 8 and 7 some stations will lose broadcasts caused by the urban density and test spots distance. In next Fig 6 we depict the most utilized gateway from 131 broadcasts among different locations and SF modes. Station A takes the majority of listening ability with 103 of them, while the rest are received by the other stations and A simultaneously.

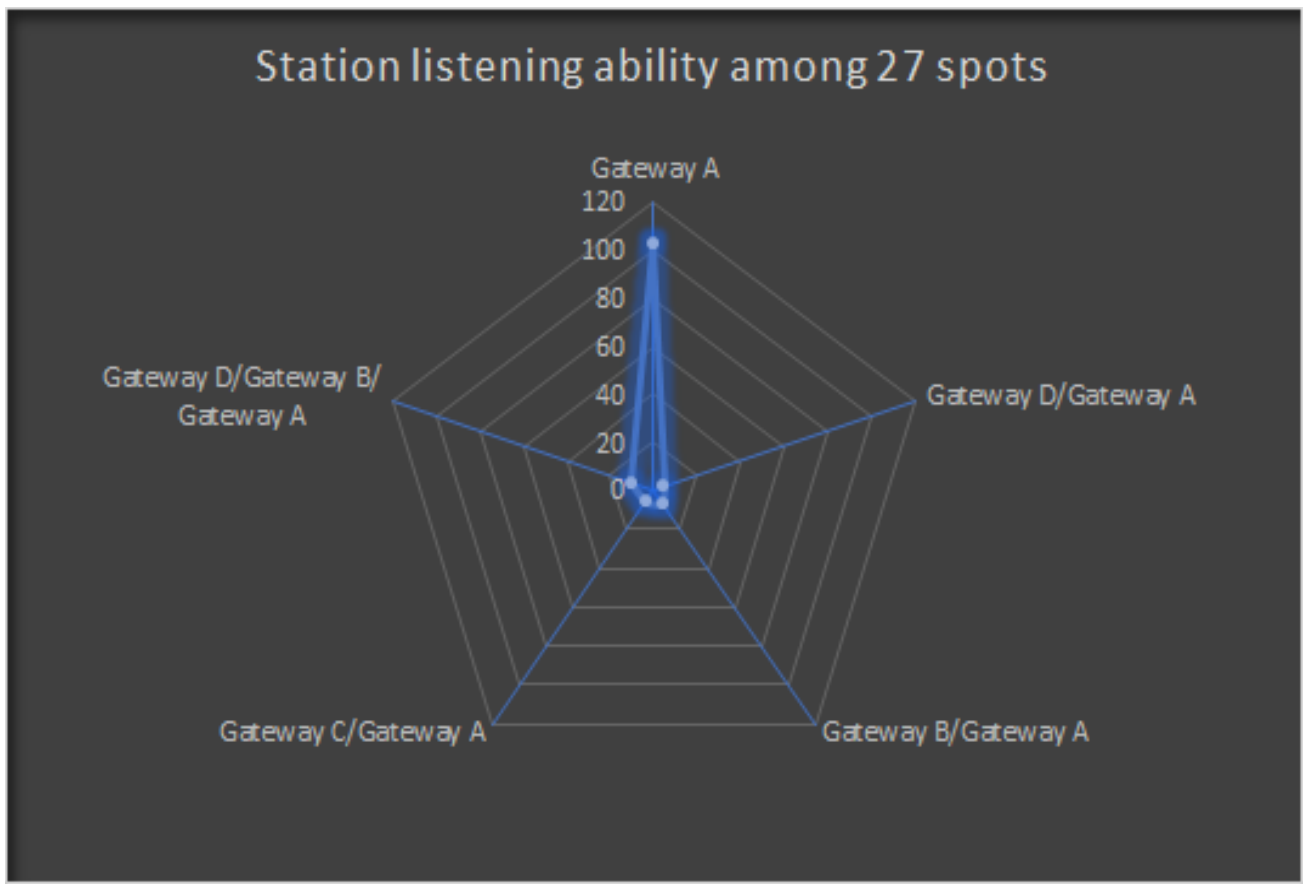

**Figure 6.** Base station activity among all broadcasts.

This concludes that a higher elevated GS at 1426 ft ASL plays a considerable role to intercept most of the end node broadcasts [32]. The portion of this field experiment was conducted with the end node operating inside a car interior without an outside mounted antenna. Relatively this affects the receiving ability at various SF modes and RSSI and SNR level inaccuracies on the end results.

### 3.1.3 Rural communications

Another area to commit radio link quality evaluation derives on the red-circled area shown in Fig 4 and the red push pins. The defined rural area extends on a radius of 15,05 km with the Gateway A as the center point and an approximate effective area of 350 km2. On those A-I test spots we accomplished 21 successful payload transmissions at various modes with a mean SF usage of 11,04 and upwards. It proved that SF12 mode can successfully be used for the 100% of all broadcasts, while modes 11, 10 and 9 offer partial potential. In SF modes 8 and 7 no gateway had the ability to listen to the end node with the percentages to be shown on the following Fig 7.

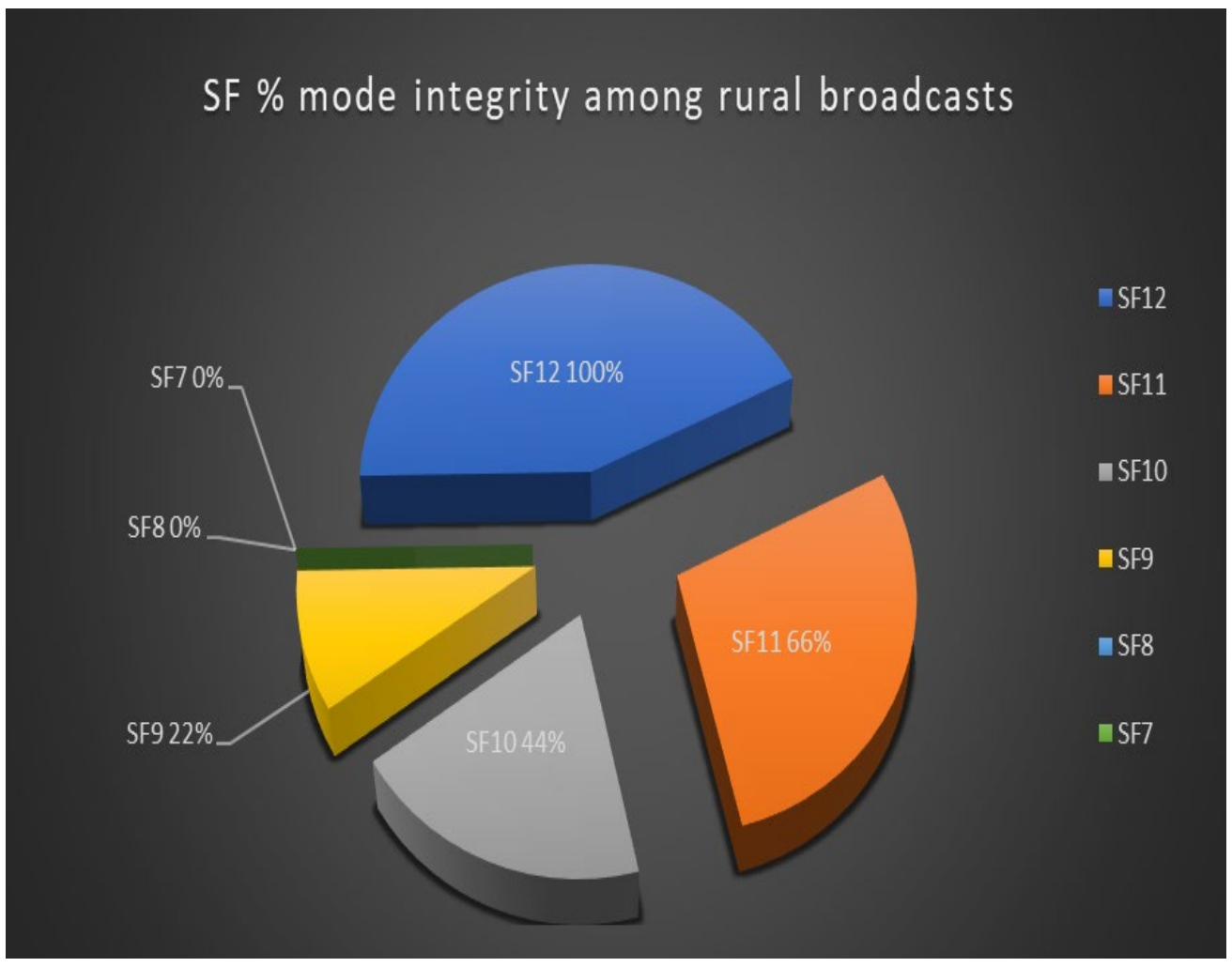

**Figure 7.** Spreading factor ratio for rural areas with percentages on total broadcasts.

The mean performance for RSSI and SNR in rural areas is in the range of (**-117dBm and -5.45dBm**). RSSI holds a negative value and is measured in dBm with values closer to 0 depicting signal robustness. High SF levels pose higher receiver sensitivity than lower ones, hence further the distance they can cover. LoRa SNR span between -20 and +25 dBm where anything > 0 means that the RSSI values are above the noise floor and the signal is less corrupted from interference. A link can be considered good when the RSSI > -115dBm and SNR > -7dBm. The end nodes we evaluated are the **RAK811** and **Ai-Thinker RA-08H** with sensitivities of -130dBm and -138dBm respectively. Table 5 shows radio metrics among all evaluated areas.

Table 5. Radio metrics among different locations

| | Solid SF potential (%) | Partial SF potential (%) | avg RSSI/SNR dBm |
|---|---|---|---|
| Urban-Suburban | 12, 11, 10 (100%) | 9 (70%), 8 (66%), 7 (48%) | -110/1.44 |
| Rural | 12 (100%) | 11 (66%), 10 (44%), 9 (22%), 8 & 7 modes (0%) | -117/-5.45 |

## 3.2 Satellite

Rounding the communications spectrum, no protocol neither a terrestrial LPWAN is immune for lack of infrastructure or infinite range [33]. For example, a sensor node meant to be serving locations such as forests, uninhabited pieces of land or undeveloped countries will face challenges for data transmission and reception. Researchers at [34] are emphasizing the broadened IoT connectivity prospect via Satellites. Low Earth Orbit (LEO) constellations can integrate data relay functions over terrestrial ones. Satellites are classified by their mass and orbit on low, medium and Geosynchronous Equatorial Orbit (GEO) classes. The survey analyzes the challenges and communication protocols with Direct-to-Satellite (DtS-IoT) topology. They suggest Lora and NB-IoT are among the most widely spread protocols summarizing them as a viable solution with multiple use cases. Notable additions are the LEO CubeSats in which academic institutions are developing them for various experimentation and communication purposes and made on mainstream and commercial off the shelf components. Manufacturers and project base entities that offer IoT space and mission expertise are FOSSA, Newspace Systems, Lacuna, Sateliot, Libre Space Foundation and SatNOGS Network. Another study on [35] specifically is exploring the signal propagation integrity at 433 MHz frequency spectrum and examines the LEO satellite ability to serve as a LoRaWAN gateway from ground data collection nodes in areas lacking internet connectivity. Although this is a theoretical study between the Slant range of an object and the Free Space Path Loss (FSPL) calculation, they achieved a theorem of the signal reach to space. Their experimentations consist of a RAK2245 LoRaWAN gateway, SX1278 end nodes and artificial attenuators to simulate the path loss. So, each attenuation level was examined for RSSI and SNR values. This proved that a transmitted signal towards space can reach a distance of 2700 km, much further of a CubeSat at an altitude of 550 km. They add that this experimentation needs additional work for the complete uplink and downlink study of the payload to the ground stations. In [36] they reinforce the notion for Extreme Edge Computing with Sat-IoT applications.

### 3.2.1 LPWAN satellite communications and use cases

**Environmental and remote area monitoring:** Researchers can deploy IoT nodes in remote areas to monitor conditions such as weather patterns, water and air quality. Satellite communications enable the nodes to transmit sensor data back to a central server for analysis.
**Precision Agriculture**: Farmers can benefit with IoT technology by monitoring soil moisture levels, temperature and other phenotypic parameters in their orchards. Satellite connectivity ensures that data can be collected even in areas with poor terrestrial network coverage.
**Asset Tracking**: Companies can track their assets status such as containers or equipment, using IoT devices connectivity. Satellite communication allows for real-time tracking across vast distances, even in remote locations.

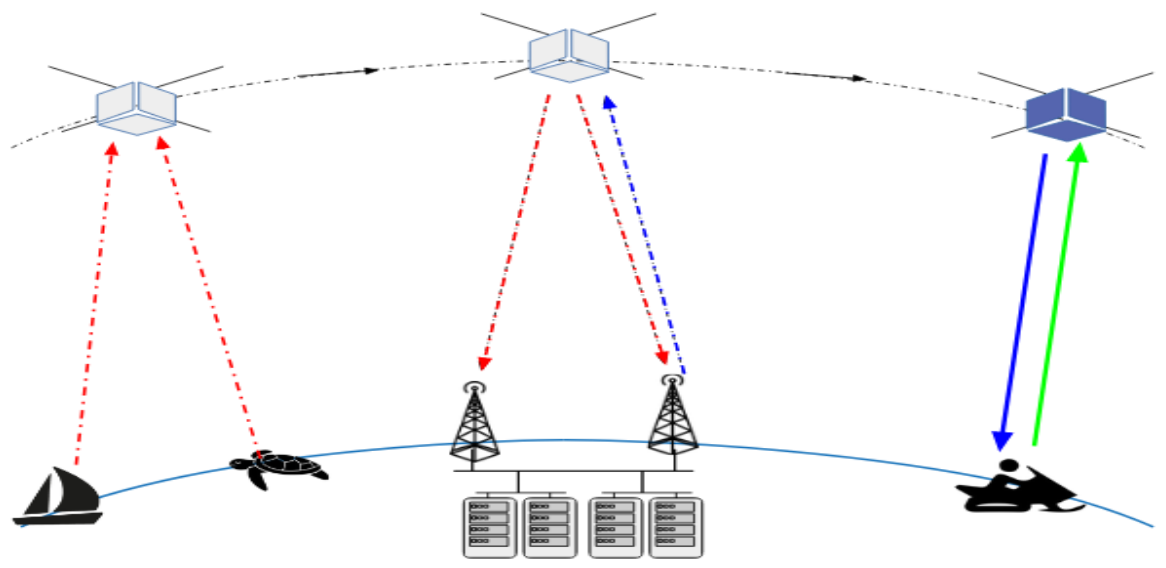

**Figure 8.** Depiction of SatIoT [37]

### 3.2.2 Time delay sensitive services

For countless IoT applications and end nodes, listening to terrestrial GS is the de facto occurrence for uplink communications. In different scenarios if conventional methods are non-existent due to lack of infrastructure, satellites can be the only communication option. A satellite has a unique term for temporal resolution also known as repetition rate, this is the orbit time interval over the same area and ranges from 14 days to 15 minutes based on the satellite type [38]. This leads to an important consideration whether a reception delay can be overlooked or is crucial for an intendent scenario. IoT Delay Tolerant Applications [39] or (DTA) are characterized by the requirement of continuous network connectivity with a tolerance delay ranging from milliseconds to several seconds. Controlling remote assets with SatIoT can be successfully utilized in various industries as this integration offer numerous benefits for environmental or agriculture monitoring which do not demand strict latency requirements and are possible to function within a tolerable delay. On the other hand, Time Delay Sensitive services or (DSAs) discerned as to those of real time monitoring significance, for example a critical infrastructure will require multiple redundancies. Delay sensitive applications for autonomous vehicles, or industrial automation will demand communication links with latency guarantees. Additionally existing upper layer protocols need to be redesigned to support these applications effectively [40]. Various studies are exploring Medium Access Control (MAC) protocols and resource allocation mechanisms for satellite aided IoT networks to meet the requirements of delay sensitive applications. They assess the conformity of current MAC and upper layer protocols for DtS-IoT communications especially in disaster management scenarios due to the sort transmit opportunity window affecting non-Geostationary satellites or LEO's with the susceptibility to Doppler effect [41]. Some of the key findings from these studies include:

- Existing MAC protocols such as Aloha and Slotted Aloha have been evaluated for satellite aided IoT networks, found to be inefficient due to high collision rates and limited access opportunities.
- Time Division Multiple Access (TDMA) has shown promise in reducing collisions and increasing efficiency for satellite aided IoT networks, particularly in scenarios where multiple nodes need to access the satellite simultaneously.
- Dynamic resource allocation mechanisms based on channel conditions and traffic patterns have been proposed to optimize resource utilization and improve overall network performance.
- Hybrid approaches combining multiple MAC protocols have been proposed to address the challenges of satellite aided IoT networks, such as the combination of TDMA and Carrier Sense Multiple Access (CSMA) to improve efficiency and reduce latency.
- Future research directions include the development of cross-layer design approaches that jointly optimize MAC, routing, and transport protocols for satellite-aided IoT networks meeting the needs of delay-sensitive applications.

### 3.2.3 Doppler effect

In a recent academic source [42], authors investigate the multiple parameters affecting the Lora DtS links regarding the Doppler effect in the LEO framework. In DtS scenarios the frequency shift is attributed by the rapid satellite movement. Physics behind it are the vehicle elevation angles in regards to the ground station, leading to reduced visibility intervals on low altitudes. The static and dynamic doppler would account for packet losses and hindered demodulation as these will introduce variability between the transmitted and received node frequency. Among these PHY frequencies, the 433MHz spectrum is preferred over the 868MHz due to interference immunity. Whether is a feasible concept apart from the technical challenges, the statistical point of view dictates a growth in active LPWAN enabled LEOs. A study in [43] sheds light on the high immunity of Lora modulation on DtS-IoT for SF modes $\leq$ 11 and bandwidth $\geq$ 62.5 kHz.

### 3.2.4 A practical attempt

To understand the DtS concepts we operated the TinyGS network [44]. Lora enabled ground stations are distributed globally and operated by ESP32 embedded boards running specialized firmware. The purpose of TinyGS is to develop communication means with satellites and flying weather probes with small and versatile devices. This can be the foundation for broadened expertise beyond the terrestrial LoRaWAN and IoT gateways towards DtS communications by the public. The device we operated is the **Heltec wireless stick V3**, connected to a dipole antenna. Both the end node and the antenna are tuned on using the 400 to 436 MHz spectrum. In the following Fig 9 we see a coverage radius of approximately 1500 km, which in theory serves an area of 7 million square kilometers from a single LEO satellite. In Fig 10 we see various LEO satellite interval transmission footprints.

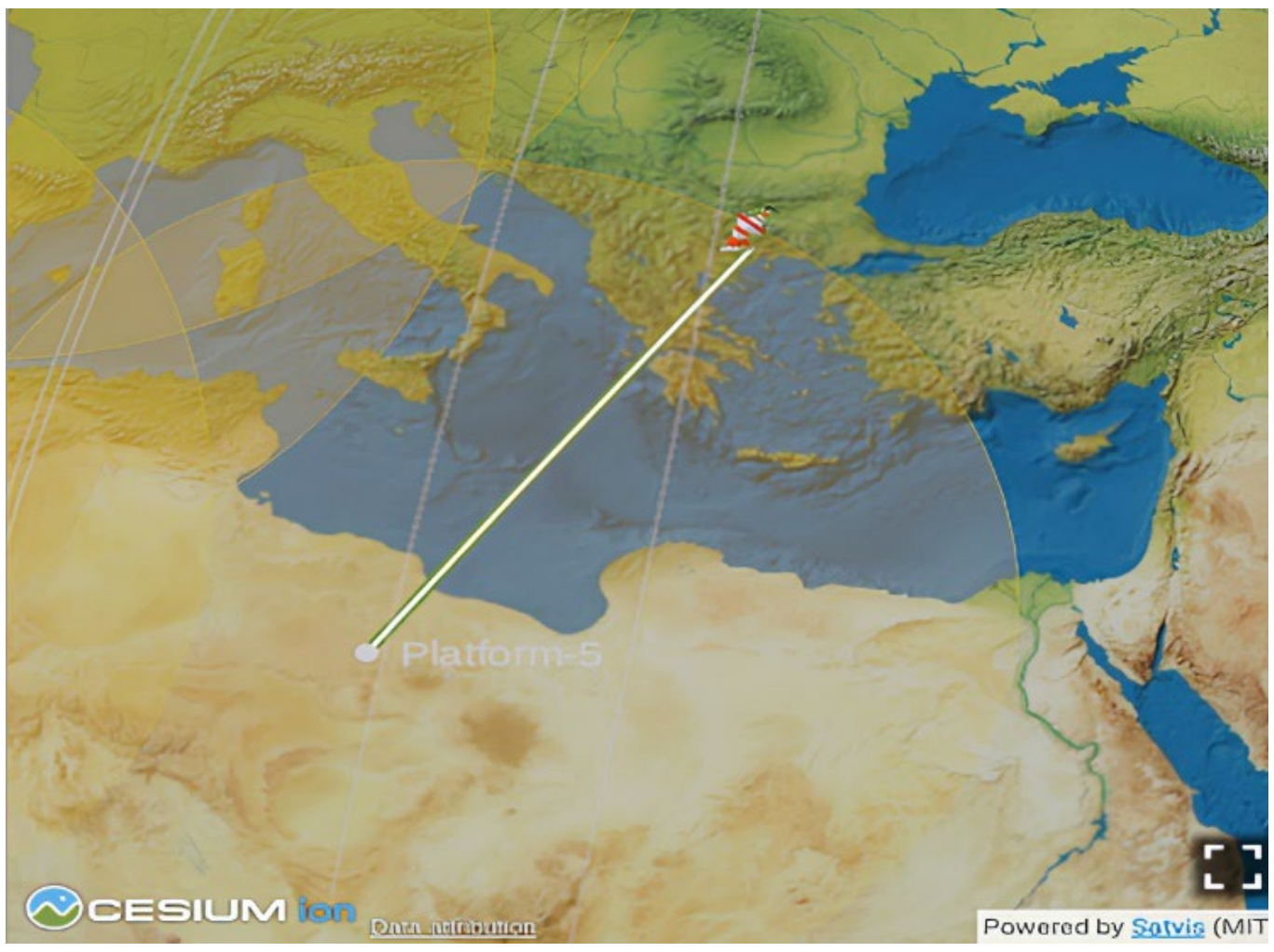

**Figure 9.** LEO position and theoretical cover radius

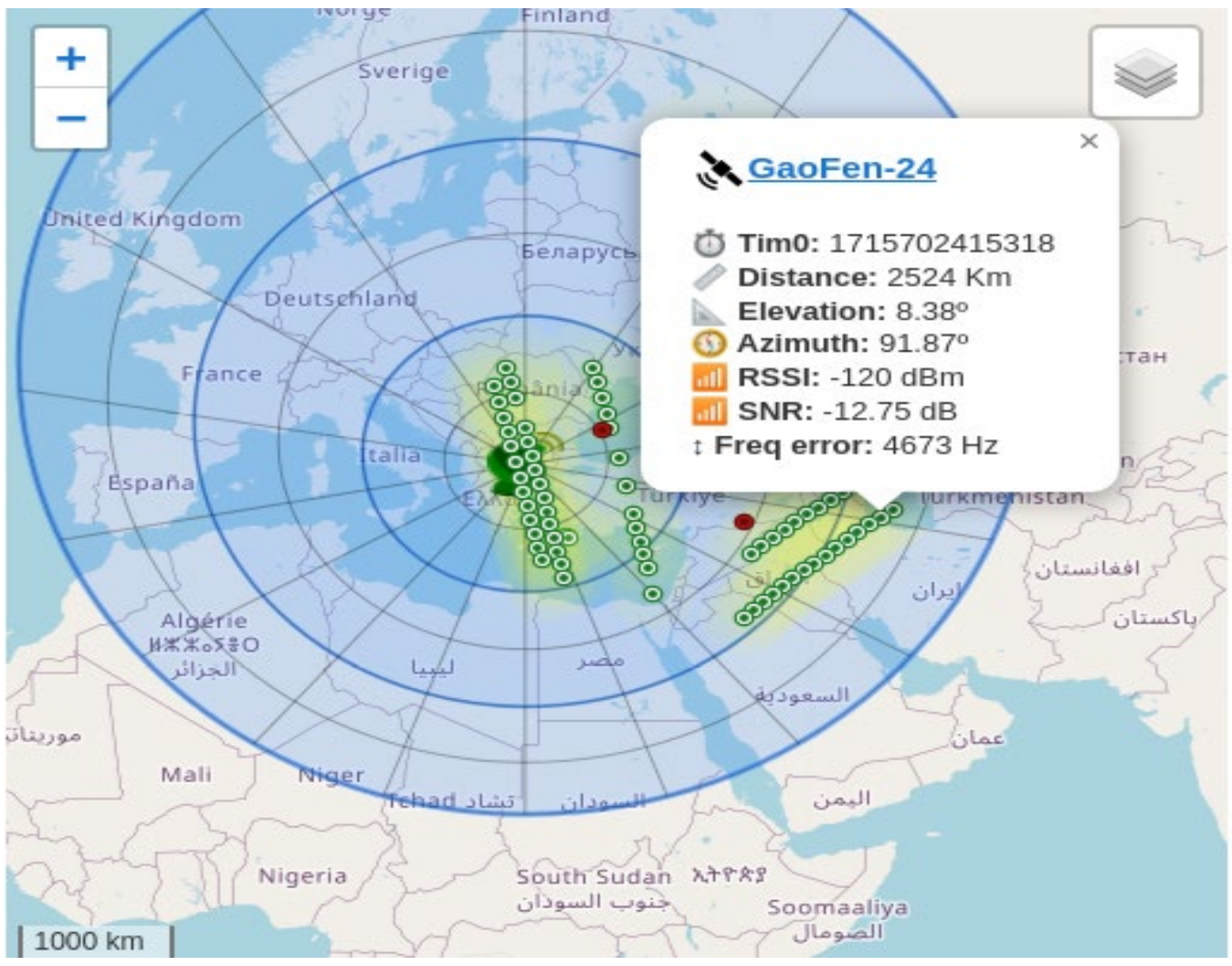

**Figure 10.** SatIoT Tx towards our ground station

From the GaoFen-24 LEO recurrent transmissions in Fig 10, a track of the first payload was picked above northern Saudi Arabia with the last on the southern part of Caspian Sea. The covered distance of 1360 km with a LEO speed of 7,55 km/sec gave 90 km Tx intervals with a 3-minute window of opportunity for Tx and Rx.

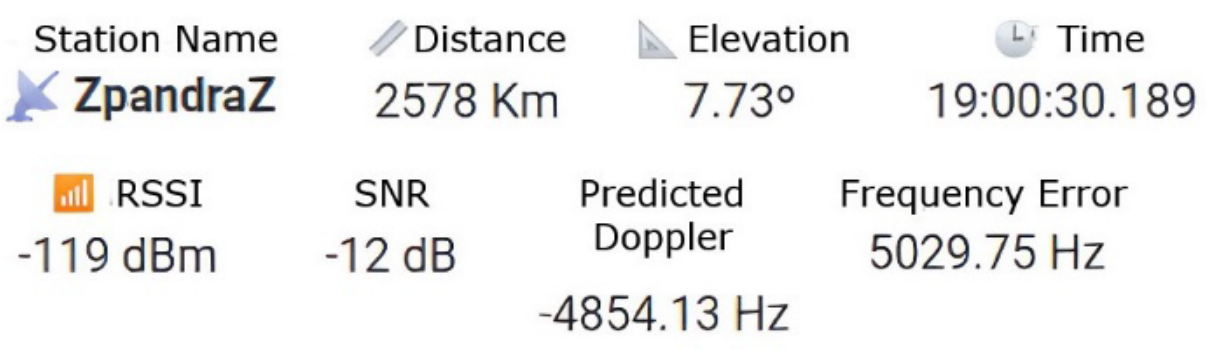

**Figure 11.** Ground station statistics

The information on Fig 11 above depicts the frequency error as an offset between the expected to the altered received packet frequency. Predicted doppler is the calculated frequency drift due the elevation and speed of the satellite. Although a very important aspect here is the absence of an uplink capability from a TinyGS node to a LEO and downlink the way the traditional LoRaWAN gateway listens. The lack of full duplex communication from the overhead LEO gateways hinders the expansion scheme for public experimentations. In conclusion, Satellite IoT and LoRaWAN technologies play a crucial role in the advancement of EC by extending connectivity to remote and inaccessible areas.

## 4. Computer Vision

Applied CV in robotics and other disciplines enable image acquisition to directly trigger actuation and provide recommendations in various scenarios. Fig 12 shows a typical ML object detection algorithm flowchart.

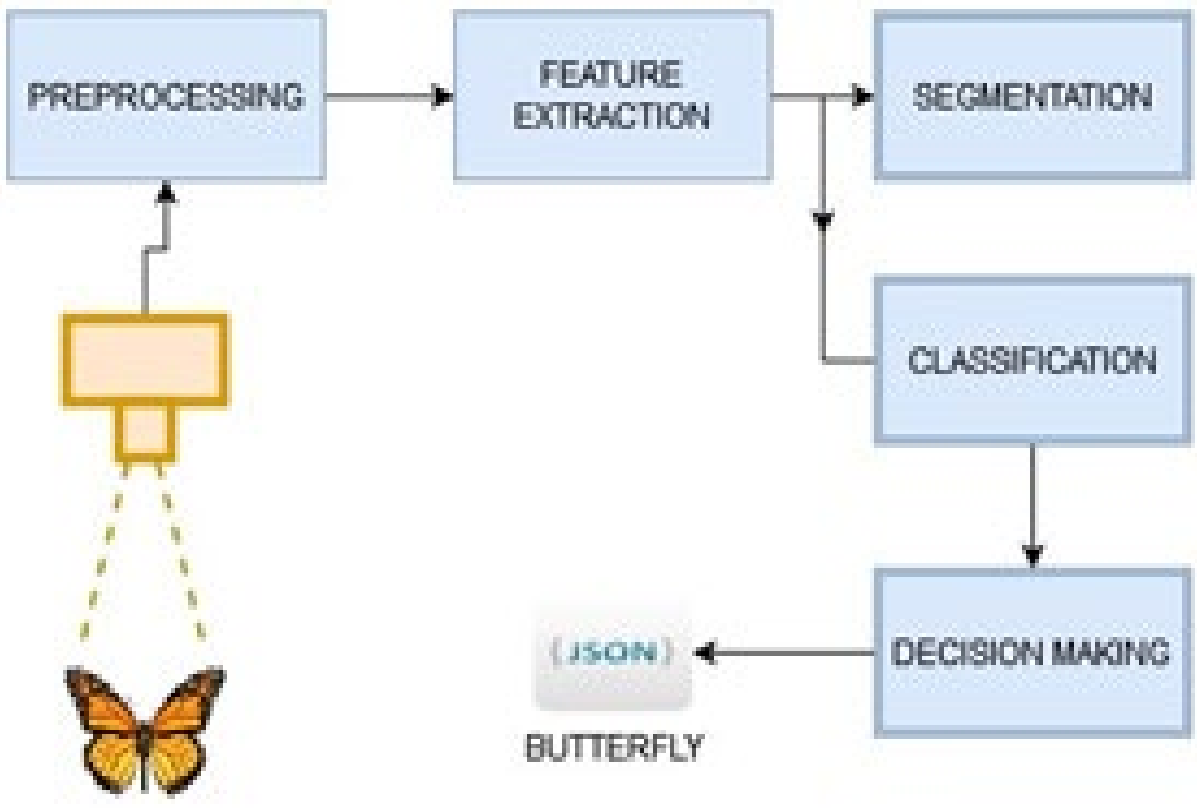

**Figure 12.** A typical CV system employing detection and classification

The aim is to detect spatial patterns, defects, contaminants and other cases using various recognition algorithms. The term CV in [45] is examining 2 definitions. First the biological scope of an interdisciplinary science that aims at computational models influenced from human visual perception and second the engineering scope of aids that perform or outperform the human vision. State of the art CV systems extract their data in 2D or 3D based on single or multiple cameras, streaming sources or data acquisition from numerical and symbolic representations. Collectively CV plays a crucial role in tasks such as object recognition, image registration, and visual tracking. Researchers in [46] proposed an autonomous Edge architecture CV system for precision agriculture. Farm biotic stress is calculated as a matter of crop yield reduction with further attention on the continuous detection of pests. They are utilizing 3 different ML classifiers such as MobileNetV2, LeNet and VGG16 to process captured images of insects inside pheromone traps in the heart of an orchard. Author's prime direction is to ensure that the EC implementation operating reliable and unattended. Second, if these ML algorithms are the right candidates for the limited resource embedded devices and third, if the proposition has viable operating characteristics with solar energy. They discuss that manual analysis from humans on pest counting and recognition taken out of digital images can be slow and error prone and the use of Convolutional Neural Network (CNN) utilized in Edge topology can overcome and normalize these limitations. In the context of Single Shot Detectors (SSD) and towards their performance bias over accuracy, in [47] authors developing a fast reaction badminton playing robot. Ball size and high-speed shuttlecock can be a prime example for a CV system to handle, especially on an embedded device. They assume that deep Yolo based network suffers from inadequate spatial information on the deeper layers and proposing the need of increased receptive fields from first layers with the use of appropriate kernels. Thus, they reduce the number of parameters without increasing the computational cost. The proposed lightweight RFSOD was tested on a Jetson Nano and achieved 30fps by the reused feature maps.

## 4.1 CV Datasets, Annotation and Camera

A dedicated dataset is a collection of images or videos that will train and test a CV algorithm performance. Popular CV datasets are the MNIST, CIFAR-10 and ImageNet. For reference the COCO dataset contains 80 classes and 1.5 million annotated objects. Our hybrid dataset includes images from, The Apple Benchmark [48], the Minne Apple [49] and pictures taken on a local apple orchard. The preprocessing was conducted using 14 mini-batches, considering pose, time and lighting characteristics, picture exposures and variations among apples. The sum consists of 302 apple images and 27 additional blanks / negatives. These negatives include lemons, pomegranates and tomatoes, items that resemble apples in detection due to color and shape. With this approach the 329 images, 2 class dataset generated 49809 annotated objects with a share of 71.74% on **apples** and the rest 28.26% for **bad apples**. The ground truth objects were density based as they could differ per image (11 minimum to 817 maximum). The dataset size is relative. For example, training a network that aims to recognize objects in a controlled environment with a fixed mounted camera will require a moderate amount of annotated data. On the contrary medium and high variable environments such as plantations are considered those which will require a bigger dataset of a few hundreds to thousands of image samples. Also, to reduce the algorithm bias on certain classes, homogenization and grouping should occur. Further annotation is the bounding boxes creation with their respective coordinates and associated classes in Fig 13.

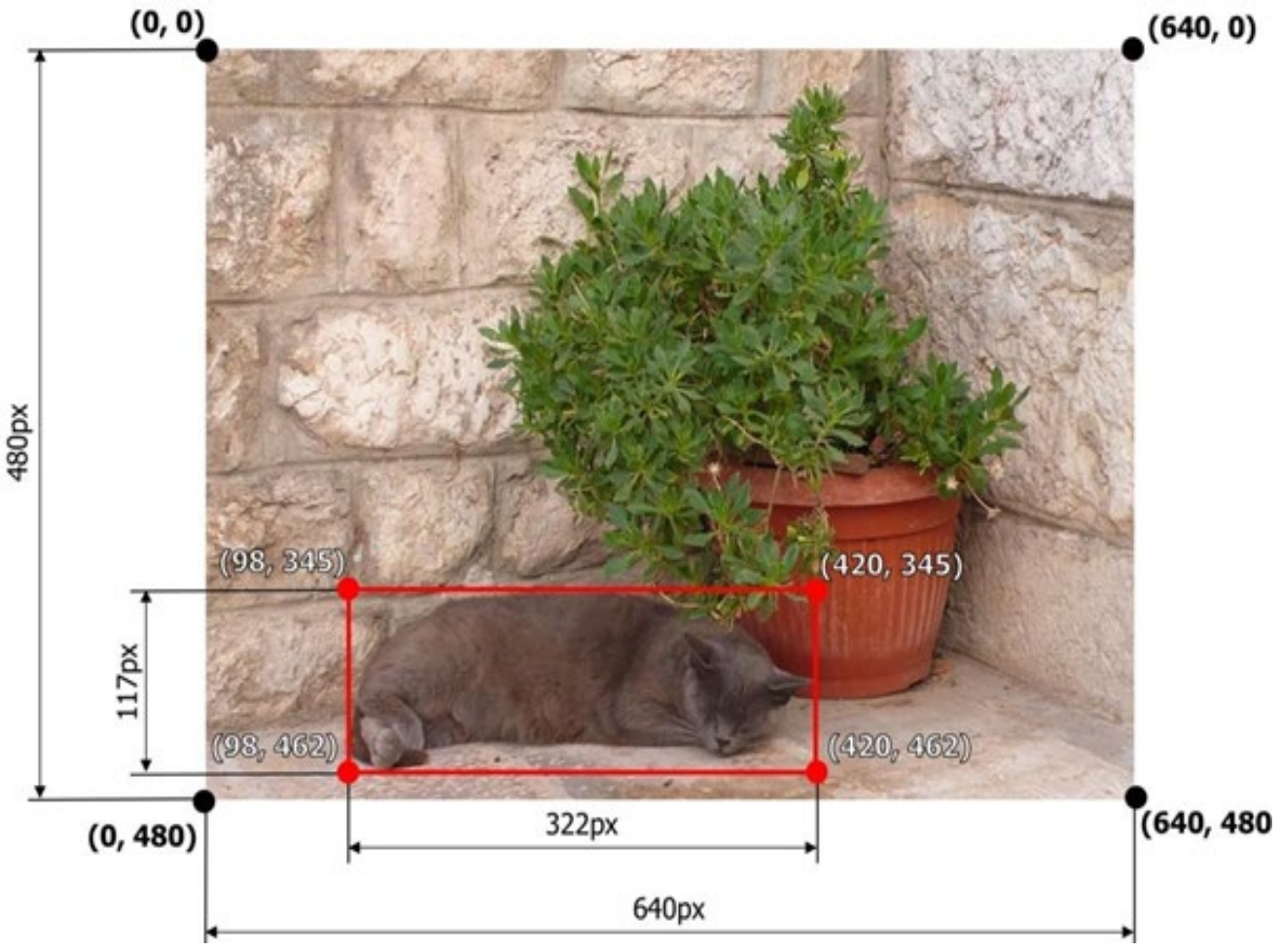

**Figure 13.** Annotation of an object, Yolo requires the center x, y pixel values. [50]

These tasks were evaluated by LabelImg [51] and CVAT image annotation tool [52], in Fig 14.

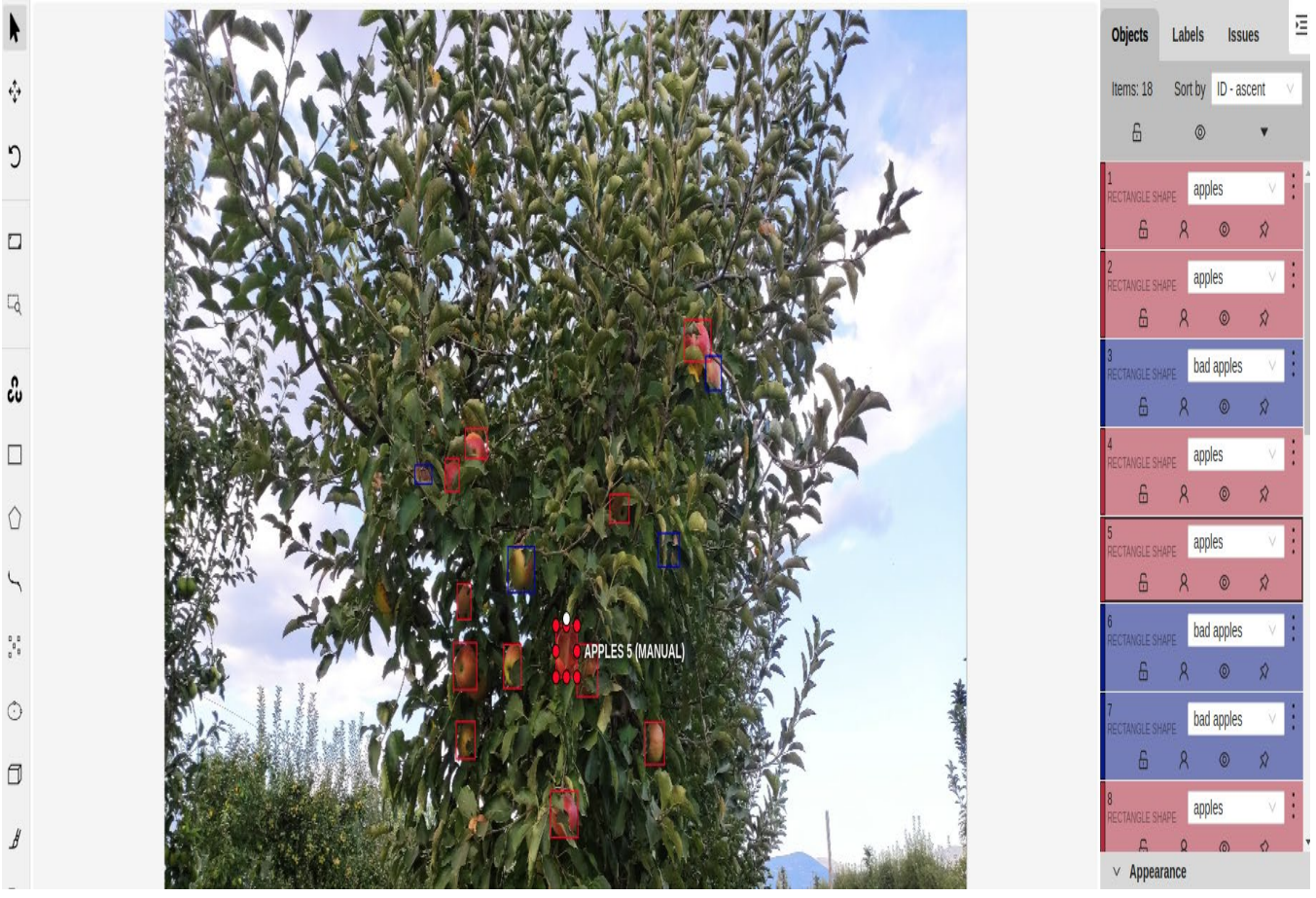

**Figure 14.** Annotation and class registration

Other types of algorithms utilize different approaches and no local annotations. The examples in Fig 15 ,16 and 17 show the different masking processes where image classification is implemented by various pre-processing techniques. PlantCV, a phenotyping library incorporates Gaussian blur, ROI (Region of Interest) and object analysis to extract leaf disease attributes. The key deriving features from the Kaggle Grapevine Disease Dataset are utilized to train a CNN that consists of 4 classes: Black Rot, ESCA, Leaf Blight and Healthy grape leaves [53].

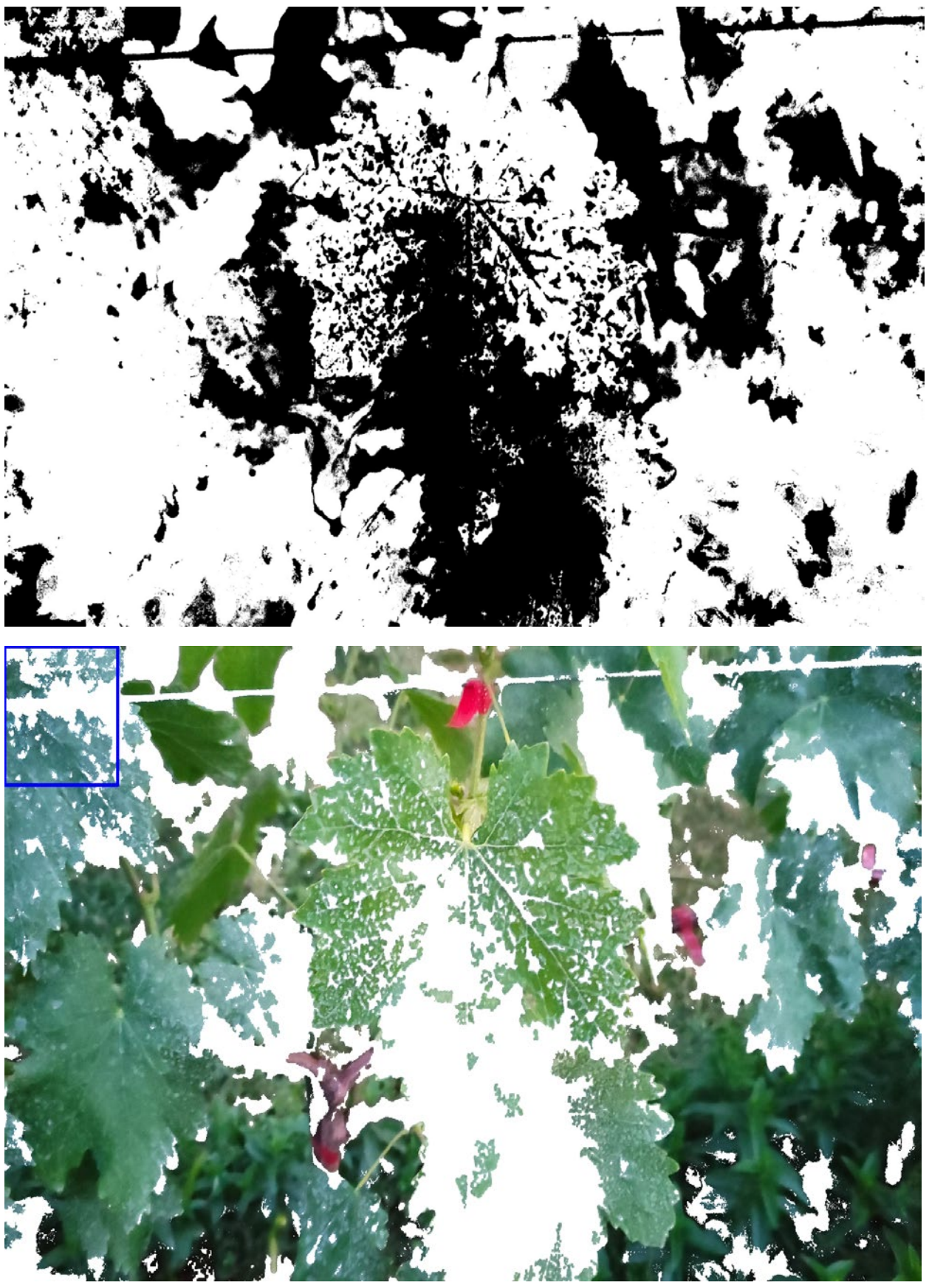

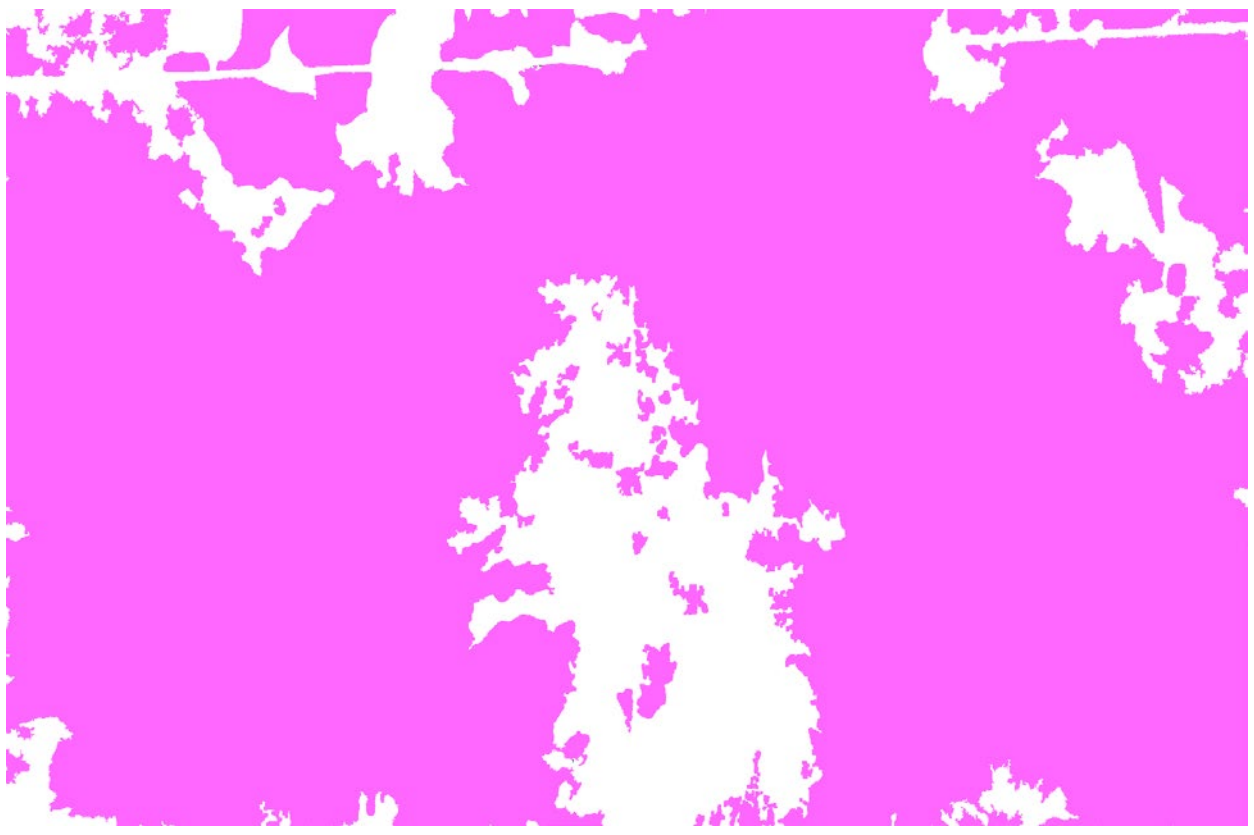

**Figures 15. 16. 17.** respectively on Gaussian Blur, ROI and Object Analysis Masks on grape leaves.

The camera as a fundamental element of any CV system has different form factors. A close distance between the camera and the subject requires different resolution to a hoovering UAV above an orchard, as more effective pixels and lens physics will help the identification and overall spatial perception. A review from [54] discussing yield prediction and fruit estimation methods for precision agriculture on different optics implemented in computer and machine vision. In our example currently, a higher resolution would identify smaller objects with the cost of higher inference time. We assessed the OV5647 5MP for Raspberry Pi4 and the Sony IMX219 8MP sensor for the Jetson Nano.

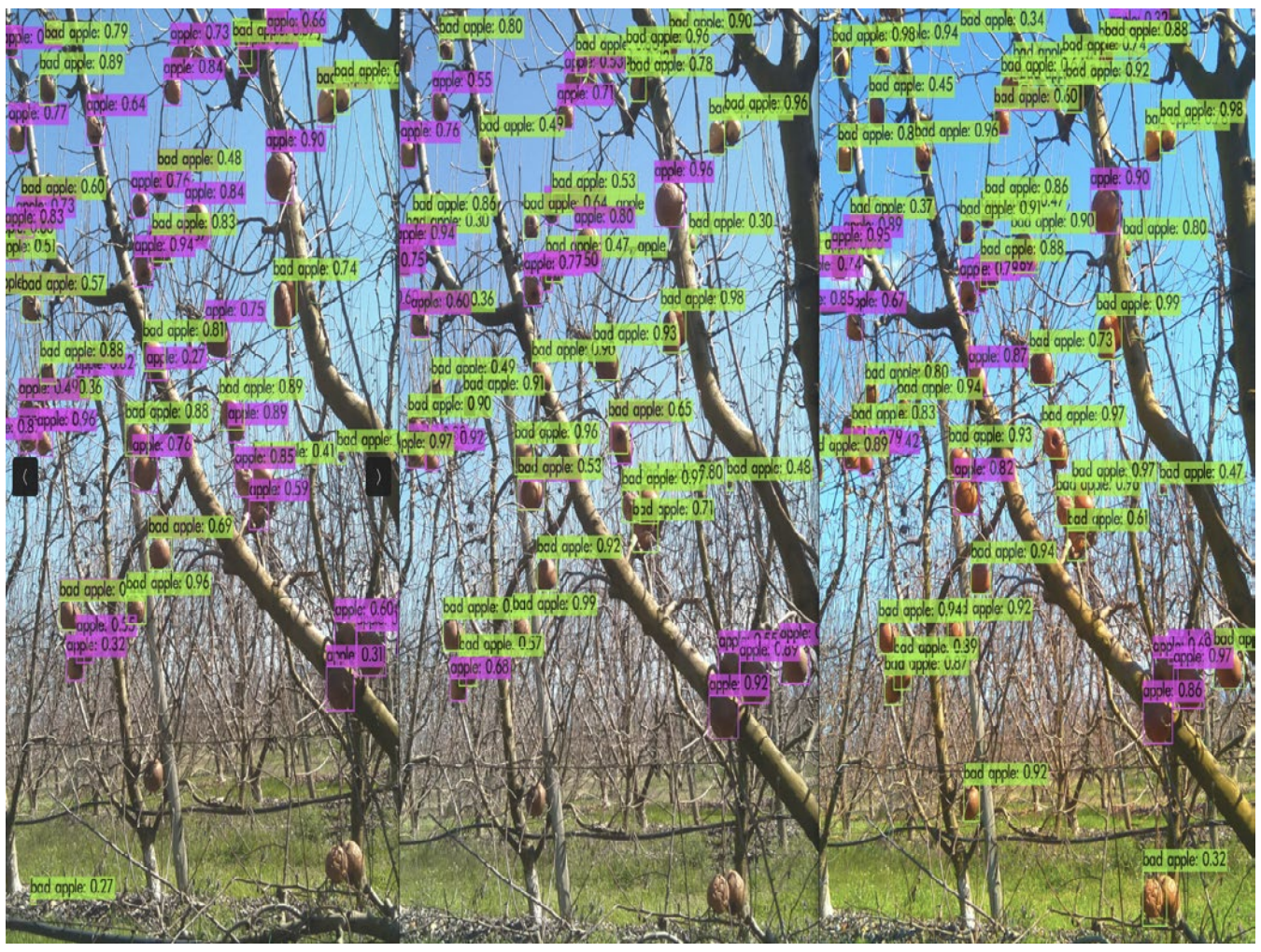

**Figure 18.** How image saturation, brightness and exposure levels affect the detection

By altering saturation, brightness or exposure levels we can emulate different optic sensors behavior. Fig 18 above shows the three subsequent image snaps of the OV5647 sensor and the varying detection results. To fully comprehend this in the next Fig 19 we used a IMX682 16MP sensor on the same angle with a slightly different distance. As such CV applications based on uncontrollable environments will require conditional sensor mounting point and image normalization techniques as angle distance, light and sensor variation will lead to inconsistent detections.

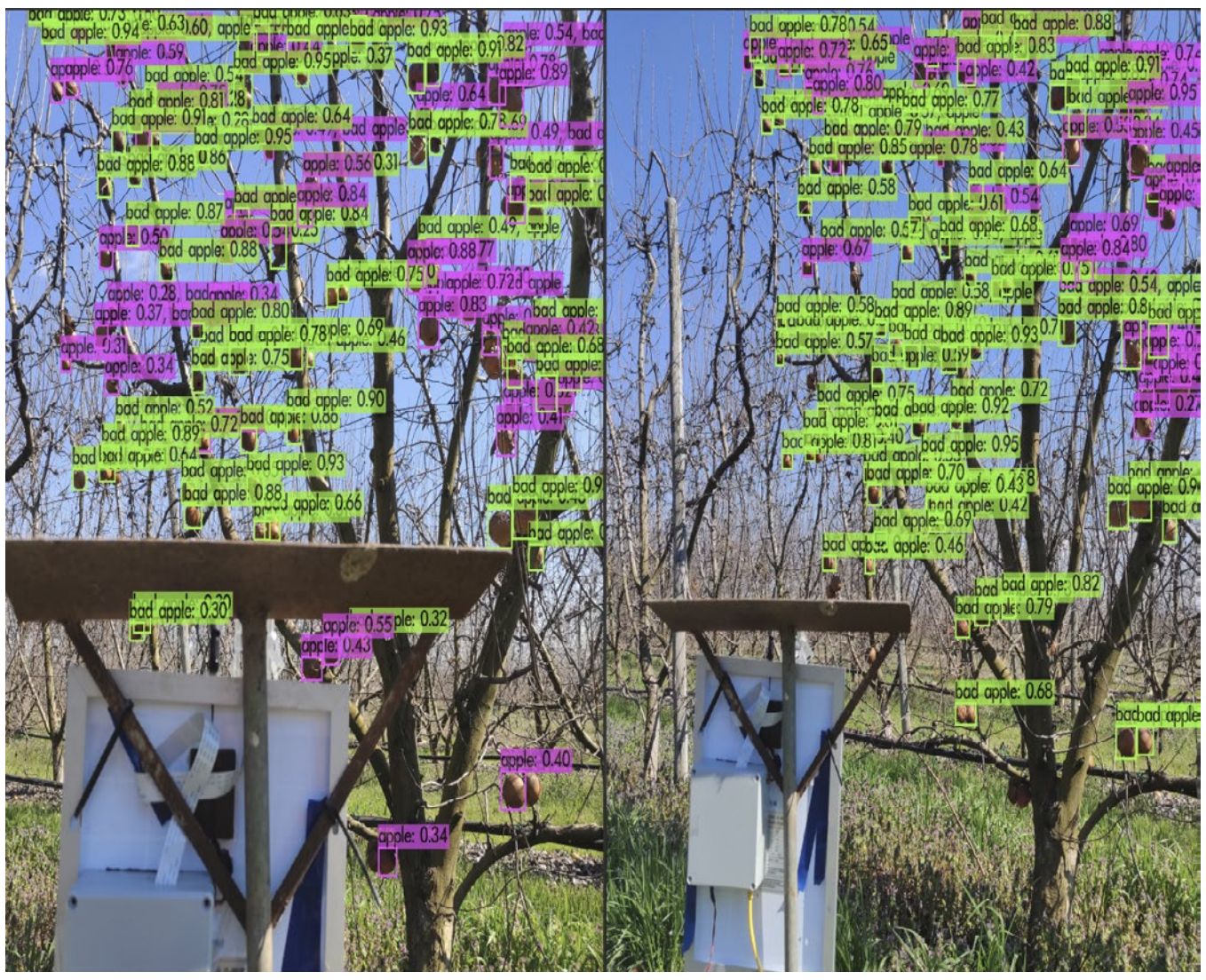

**Figure 19.** Different angles and distance.

## 4.2 Performance comparison between 2 SBC platforms and Energy Drawn

Raspberry Pi model 4B is the core platform and Darknet is the experimental framework for YoloV4 SSD. With that said the inference intensive tasks are solely use CPU time. To this extent we investigate the possibilities and differences with a Jetson Nano 2gb version featuring a GPU for faster parallel neural calculations.

- RPi has better I/O performance than the Jetson on loading weight files, due the newer architecture of the A72 Arm CPU.
- Jetson even with the reduced I/O lag, has a significant performance gain at 29.93, with RPi at 91.29sec inference time, attributed to its GPU.
- Both platforms use the Kingston Canvas Go Plus 64GB A2 performance rating.
- RPi memory usage is 183mb at idle, 1.45gb@576 and 1.6gb@608 during inference. It has a better ram management than the Jetson equivalent.
- Jetson Nano memory usage is 240mb at idle in headless mode and 450 with the GUI enabled, while at inference maxes its ram at 1.9gb and 1.1gb swap.

Table 6. Performance comparison in Yolo v4 inference between 3 different platforms

| | I5$^{10th}$ 24gb ram laptop with GTX1650 +4gb VRAM gpu | Jetson Nano 2gb Ram + Swap file | Raspberry Pi 4B 4gb Ram |
|---|---|---|---|
| Start to finish script[1] | 5.39sec@608x 608[5] | 57.62sec[4], 46.64sec[4] @608x608 | 98.30 sec@ 608x608[5] 85.04sec@ 576x576[5] |
| NN loading [2] | 4.06sec | 18.84sec[4], 16.71sec[4] | 7.01sec@ 608x608[5] 6.77sec@ 576x576[5] |
| Inference time only[3] | 1.33sec@608x 608[5] | 38.78sec[4], 29.93sec[4] @608x608 [5] | 91.29sec@ 608x608[5] 78.27sec@ 576x576[5] |

*1. Elapsed time from start to end of the script comprised on the Neural Network (NN) layers loading and the actual inference time. 2. I/O loading time, less time shows a strong performance between the CPU, the storage controller, NVme or SD random disc access performance. 3. Shows the GPU neural engine performance. These metrics are usually in FPS or in sec. 4. For the Jetson Nano 2GB only we tested both GUI and headless setup. Through Secure Shell Protocol (SSH) we disabled the desktop GUI to spare crucial RAM during inference. With this there is a 22 % speed improvement. 5. Network size / input resolution.*

The following Table 7 demonstrates various subtasks such as the I/O load, inference, data parse, LoRaWAN uplink and the reception on a frontend UI. Commercial level SBC implement slow eMMC storage options which hinders their full potential.

Table 7. Subtask time factors, energy draw

| | 704/ RPi | 704/ Jetson | 608/ RPi | 608/ Jetson | 576/ RPi | 416/ RPi |
|---|---|---|---|---|---|---|
| Image snap & NN load | 7" | 61" | 7" | 17" | 7" | 6" |
| Yolo Inference | 138" | 72" | 91" | 30" | 78" | 53" |
| Data parse/Lora /UI display | 6" | 6" | 6" | 6" | 6" | 6" |
| Total Time | 151" | 139" | 104" | 53" | 91" | 65" |
| Energy Inference draw | 0.21 Wh | 0.20 Wh | 0.139 Wh | 0.083 Wh | 0.119 Wh | 0.08 Wh |
| Energy per inference ratio[1] | 1.52 | 2.77 | 1.52 | 2.76 | 1.52 | 1.51 |
| Result | Pass | Pass | Pass | Pass | ADQT[2] | MRGL[3] |

*1. Higher value shows efficiency (× $10^{-3}$). 2. ADQT Adequate 3. MRGL Marginal*

Energy draw estimations for the RPi at peak CPU time during inference is P(w) = 1.1amp x 5V equaling to 5.5W. To measure the ***Wh*** in different network configurations then **0.21Wh** would be consumed for **138″**, **0.139Wh** for **91″** and **0.119Wh** for **78″** seconds respectively. Alternatively, in the 10W Jetson Nano the power draw will be **0.2Wh** for **72″** and **0.083Wh** for **30″** seconds inference time. High I/O time for the 704-network size in Jetson is attributed from the limited 2gb ram and slow swap file. The *energy per inference ratio* label above in Table 7, assumes that high computation availability on low demand applications is a waste of resources. At the same time, calculating the power draw for LoRaWAN uplinks, instantaneous load spikes of 50-100 mAh between SF modes 7 to 12 were detected. These noticeable fluctuations in theory dictating the power demands especially on the battery operated IoT implementations. It is clear that the Jetson Nano outperforms RPi significantly in inference as it is 3 times faster (30″ vs 91″ at 608×608 network size) and in energy almost 1.67 times lower. All along it means that latency tolerant implementations such as precision agriculture or infrastructure inspection can remain viable on either platform, though with safety critical cases such as traffic incident detection and others, GPU enabled Edge nodes with latency mitigation analysis is imposed.

## 4.3 CV Metrics and field behavior

Certain metrics are employed in data sciences to assess the trained algorithm. *Precision* defines the correct predictions as a ratio of true positives (TP) divided by the combined TP and false positives (FP), *Recall* is the proportion of TP to the total sum of TP and false negatives (FN). Both can be raised proportionally in an optimal model. The concept of IoU in Fig 20 represents the intersection area of the predicted bounding box (red) over the union which corresponds to the actual ground truth (green). With a 0.5 threshold our TP derives after an overlap of half the ground truth and above as a confidence factor. If we lower this threshold then more samples will be identified as TP improving recall and precision scores.

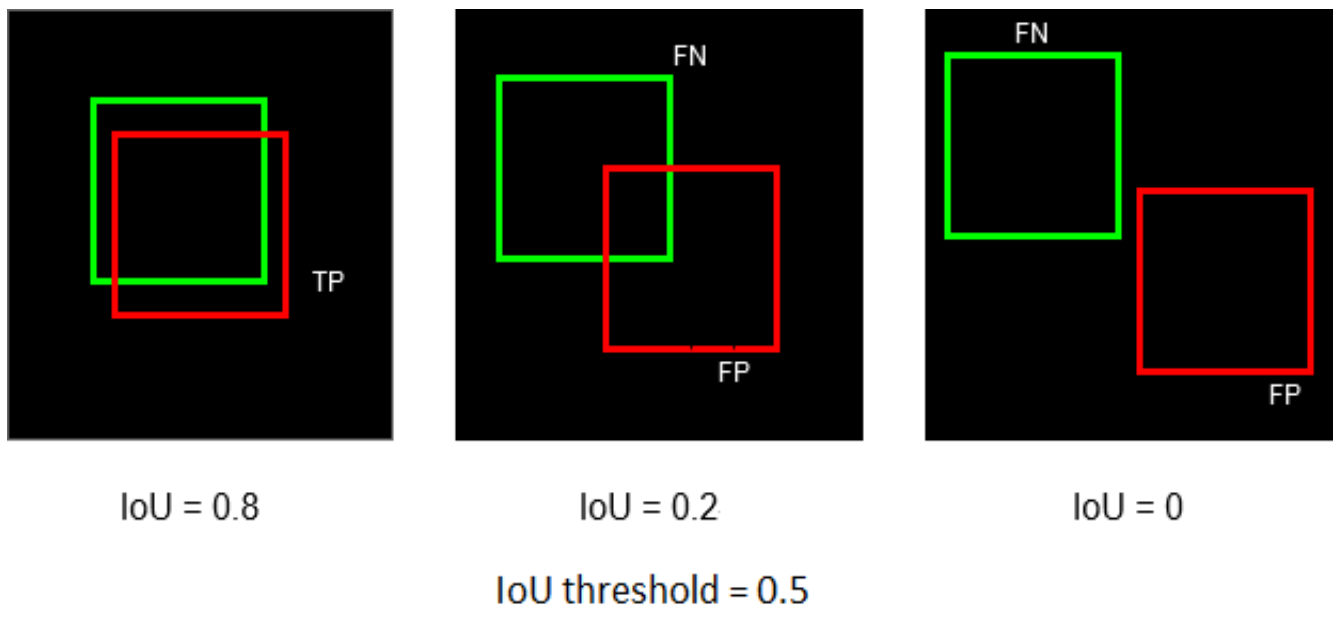

**Figure 20.** The concept of IoU

YoloV4 will be validated on a 20% of unseen data of the original dataset. In the following Table 8 the best achieved batch identified as the **3700@329** with **0.76** precision, **0.65** recall and **0.70** *f1* value. The latter shows the robustness against other training iterations as a harmonic mean between precision and recall. In simple terms the lower it gets is indicative of model imbalance among the basic CV metrics.

Table 8. Final training iteration trials on Precision, Recall and f1 metrics.

| Iterations #images 704x704[1] | Precision | Recall | f1 | AP good apples | mAP@0.5 |
|---|---|---|---|---|---|
| 2[2]@329[3] | 0.64 | 0.69 | 0.67 | 73.95% | 62.21% |
| 3[2]@329[3] | 0.77 | 0.57 | 0.65 | 75.01% | 63.53% |
| 3,7[2]@329[3] | 0.76 | 0.65 | 0.70 | 74.23% | 65.52% |
| 4[2]@329[3] | 0.79 | 0.56 | 0.65 | 74.42% | 62.76% |

*1. Network Size. 2. Training Iterations * 1000 / Max Batches. 3. Dataset size*

To homogenise both metrics a precision and recall curve at varying IoU confidence thresholds is employed. The following action will plot all *(x,y)* metric values towards an optimal point. Fig 21 graph depicts the confidence thresholds starting at 0.13 on the left edge of the orange curve in 0.01 increments up to 0.8 on the right edge. This reveals that the Euclidean distance from (1,1) towards the curve intersecting the optimal threshold. A fairly robust model that operates in stable environment conditions would expect minimal class imbalances hence the precision and recall curve should be closer to (1,1). Exact coordinates for Precision (x) = 0.7 and Recall (y) = 0.7 intersects the optimal threshold of (z) = 0.15, also in 3-dimensional depiction in Fig 22. CV metrics balance are of particular importance and relevant in the IoT scenarios were transmitting fewer but reliable positives is preferable than to send large volumes of uncertain data over constrained networks. The findings highlight the feasibility but underlining the need for a larger annotated dataset, the proportionality we discussed in the **4.1** section.

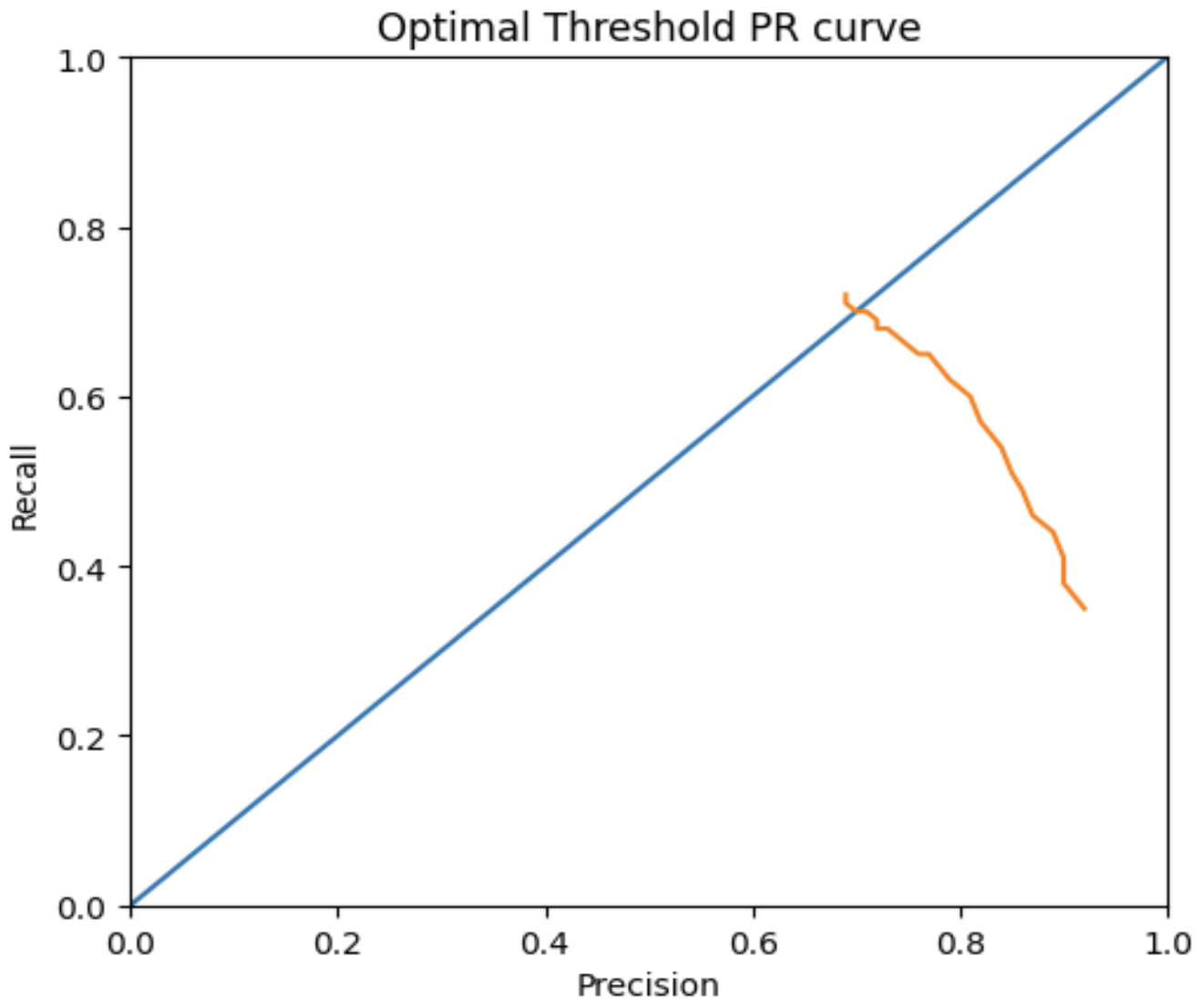

**Figure 21.** 2d conf_thresh

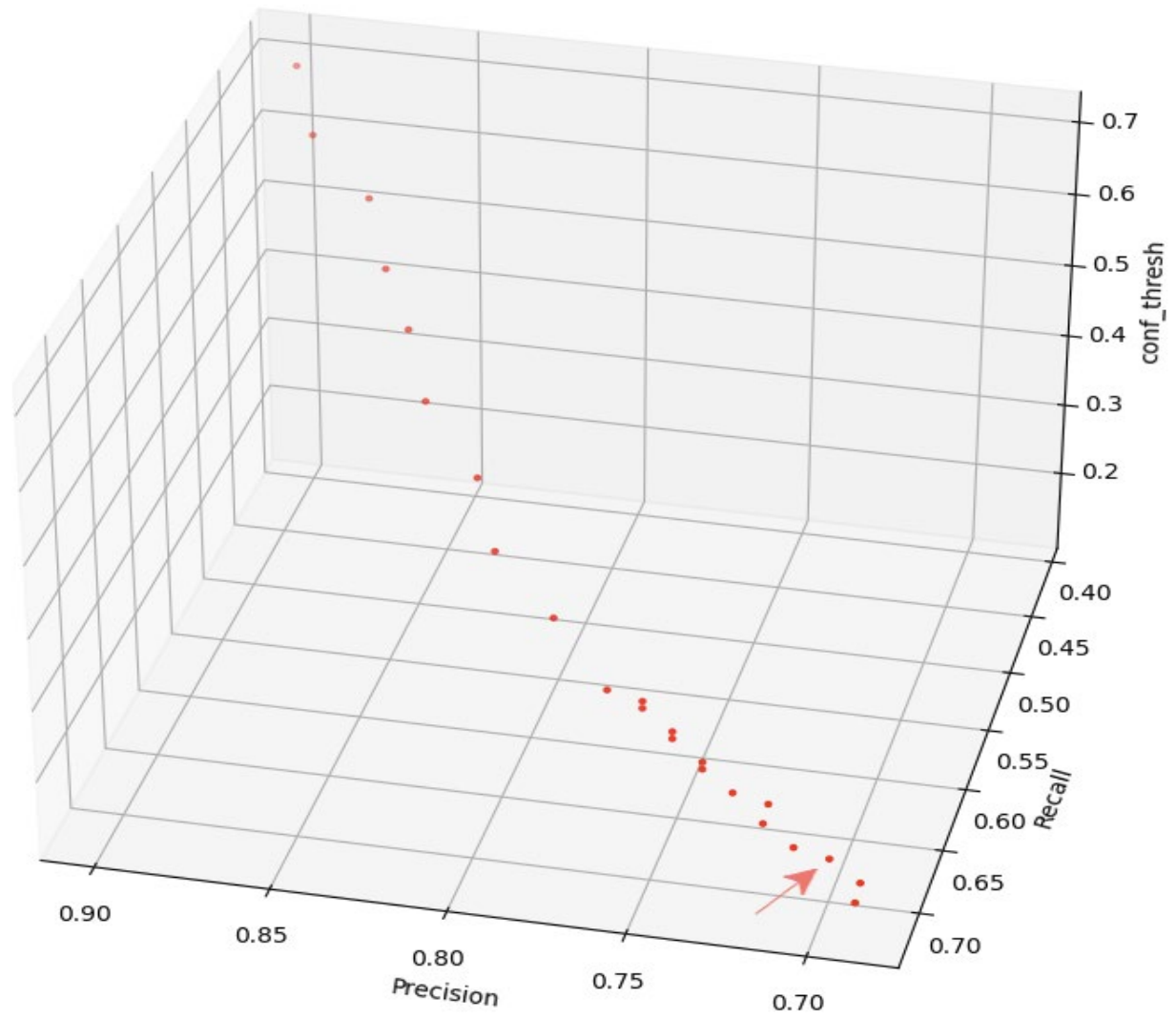

**Figure 22.** 3d conf_thresh

Based on where our model currently operates in, setting the IoU confidence threshold will cope with crop occlusions from tree branches and leaves. In this case FP objects become TP due to this predefined threshold and detect actual crops, Fig 23.

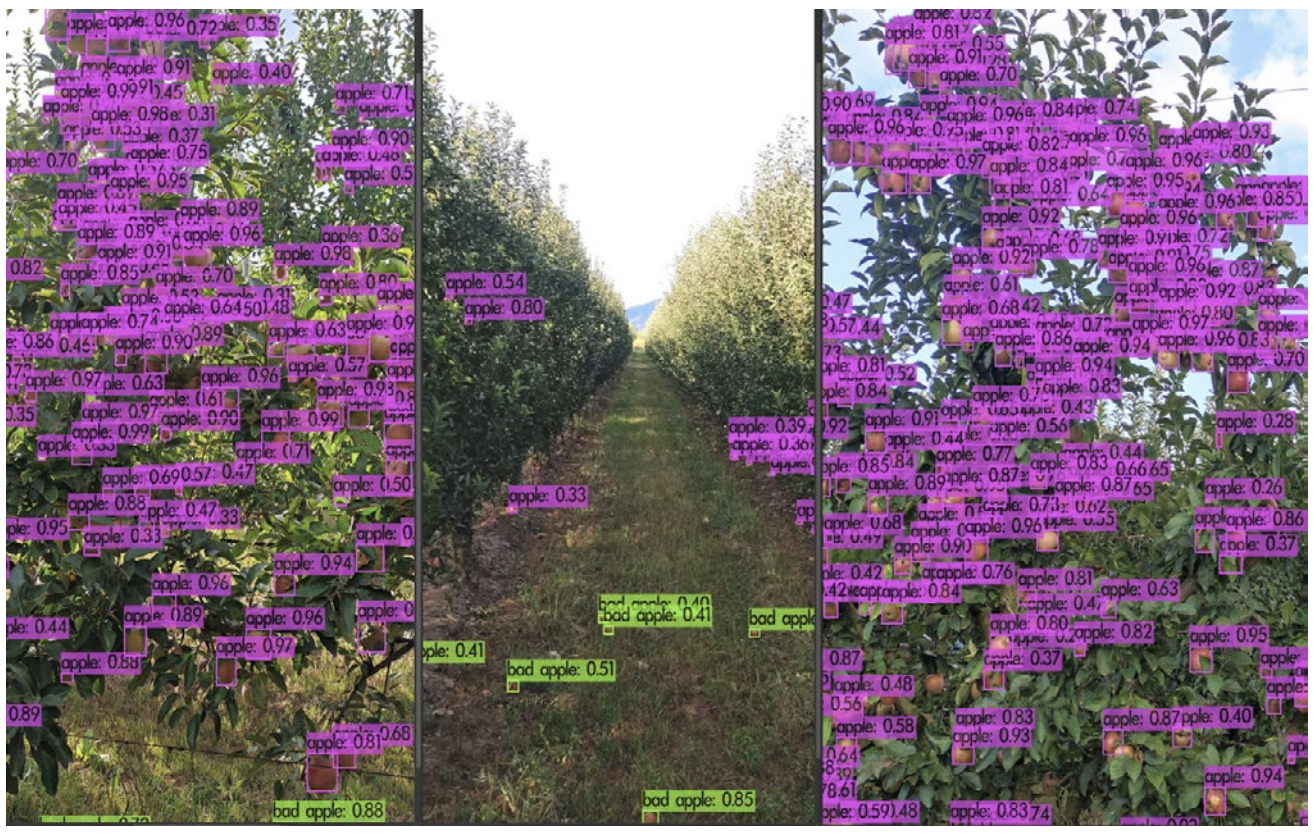

**Figure 23.** How the trained algorithm detects good and bad apples on frames.

Recent papers resonating our work focus while emphasizing the relevance of augmented CV within the IoT contexts. In [55], a modality fusion vision transformer is proposed for collaborative hyperspectral and LiDAR classification, demonstrating that multimodal architectures are achieving high accuracy. With further efficiency improvements the principle can align with the EC based fusion strategies. In [56], the RSEE framework is introduced to jointly optimize video resolution selection and conditional early exiting validated in a Jetson Nano SBC. The significance of this work is founded with the reduced computational cost while maintaining recognition performance and is directly applicable to real time vision processing on the resource constrained EC deployments. Additionally, authors in [57] exploring a multimodal fusion framework integrating RFID and CV monitoring human exercise, demonstrating that a combination of visual and non-visual sensing is enhancing IoT applications where communication and energy efficiency are critical.

## 5. Discussion

Object detection models can face specific difficulties over accuracy and robustness. Factors such as data limitations, hardware or software constraints play a significant role. Specifically, the performance in regards to detection and classification will certainly be influenced by the defined strategies and dataset quality remains the backbone. Additionally, the model architecture for a specific domain is crucial. For example, SSDs have weaknesses on very small objects while fine-grained information such as plant phenotypes will require specific segmentation techniques. To ensure optimal operation, the deployed model should be analyzed, scrutinized and refined. Furthermore, a model may encounter difficulties in adapting to novel or complex user inputs emphasizing the necessity of post deployment evaluation, sensor sensitivity and the camera placement are one the dependence factors that will complement a fine-tuned CV algorithm.

For LoRaWAN, the three case studies lead to multiple conclusions. The node configuration and gateway strategic location are indices that influence data relay performance. Similarly, antenna gain is of pivotal importance. More specifically a stationary device correctly tuned, transmitted all the intended uplinks on the urban and suburban conditions of a flat provincial city - **3.1.2**. On rural communications now due to further distances the operation was similar but the ground morphology hindrance can essentially affect data relay. As we noted, higher SF that led to longer air time can relatively tackle factors such as signal absorption, refraction or lack of LOS. The experiments prove its purpose as a resilient mean for remote sensing applications where small data, distance coverage, flexibility, scalability and power consumption are the requirements.

Regarding security, EC deployments are inherently less prone on privacy risks and attacks due to the distributed nature of the devices that reduce the single point of failure and interception chances. Also, the minimal data transmission will be less desirable for DDoS attacks. For maintaining trust and resilience, robust identity authentications and different safeguard measures should be employed.

The empirical evaluation of the two commercially available SBCs reveals that low-power platforms when combined with lightweight DL models can serve as viable edge inference actors. Primarily for latency tolerant decision-making applications. Last, scalability can remain a challenge as resource-constrained edge devices may struggle to maintain responsiveness when deployed at a higher scale. Hybrid edge-cloud frameworks though with lightweight protocols can sustain the service quality.

### 5.1 Feasibility considerations

The evolution of 6G technology hitting the markets by the end of this decade will enhance connectivity and computing capabilities. European Telecommunications Standards Institute (ETSI) segregates Edge Computing as an overarching term and MEC as an evolution for mobile communications. Current dense terrestrial networks can be in favour of low latency IoT but in difficult areas and dead zones the data uplink may be problematic. The future architecture of MEC in 6G networks [58], [59] is envisioned to incorporate various wireless communication platforms. LEO satellites, High Altitude (HAP) and Low Altitude Platforms (LAP) are potential candidates for complementing terrestrial communication infrastructure. MEC integration enables the network to provide a range of services, including communication, storage, computing, and management. The versatility, flexibility, and manoeuvrability of UAVs have garnered attention in the context of UAV enabled MEC networks. This is a strong indication that the academic community and industry will dedicate resources to promote ultra-reliable and extended coverage EC propositions. Object detection and classification is a relevant area where feasibility aspects are examined. The following work from [60] comments on the visual perception differences between a human and CV systems specifically on visual illusions. Principally, human perception is a multifaceted process that involves the combination of various visual cues such as colour, texture, motion, depth and context. In contrast CV systems typically rely on the analysis of specific visual features, neglecting the rich contextual information that is inherent to human perception. This case, features the development of sophisticated CV models that can capture the visual distinctions and complexities as human experiences. The context of intelligent systems that can process and analyse multi modal information uniformly rely on the ability to extract and synthesize multidimensional data. To facilitate the decision making on these systems, an intermediate mechanism is necessary. The survey in [61] focuses on a key challenge to extract visual, textual, and other representative attributes from multiple data streams and distribute them into a common representation space known as Multimodal AI. The success is driven by the availability of large, widely usable data sets, powerful computing resources, and high-quality feature representations. However, the open research challenge is to strengthen the correlations through robust models and select the optimal fusion schemes.

### 5.1.1 Where does this feasibility study lead to?

Answering to what this feasibility study acts upon on has a multi-dimensional approach. Technologically; innovative EC systems in their ability catering many use cases on varying levels of maturity. They are characterized by their scalability and low latency characteristics where data collection is obtained from scattered IoT or imaging sensors. A useful roadmap by [62] examining large model quantization can deliver effective deployments on smart cities, autonomous vehicles, industrial automation and healthcare. In socioeconomic planning; EC can offer various services especially in rural areas and developing countries where the pervasive issue of internet access exists. Precision agriculture can be utilized efficiently with these services in hunger and undernourishment communities on the emerging issue of global food security [63]. The following publication [64] offers a comprehensive technical analysis for the economic opportunities and environmental benefits associated with the shift towards EC within the EU markets. AI and ML integration at the edge represents a significant innovation in the digital transformation that leads to the creation of new business models. The study also focuses on the importance of LPWAN and satellite networks while concluding Edge AI computation rather than in the cloud computing yields more accurate results and enhances the efficiency of AI algorithms. In the evolution phase; the transition between the current Legacy AI to the integration of Generative AI models for decentralized based entities is taking place. Google AI studio and Nvidia Metropolis Microservices are examples where the interaction among visual and contextual representation is taking place. These services are enhanced Edge to Cloud integrations, with pre-trained models on different modalities based on internet scale data to reason unseen classes. Zero Shot Object Detection (ZSOD) in [65] and Visual Language Models (VLM) represent distinct areas of AI where the explicit labelled training data is not a requirement and the integration of cross modal reasoning respectively to link visual and textual information will both enhance object detection tasks. The following work by [66] examining the cases on tuning Large Language Models (LLM) by inducting visual inputs and integrate them into a Visual Language (VILA) that can be deployed on the Edge of the Network with a Jetson Orin platform. What is certain, semiconductor manufacturers are committing into the EC sphere and develop AI Processing Units (AIPU) designed for scaled edge inference workloads, for low power mobility applications or for workstation class systems [67]. Additionally, the following paper [68] highlights how scaled Gen AI systems leveraging Edge-Cloud computing can solve current infrastructure challenges more effectively by combining local and remote computing resources (offloading) to reduce latency and handle more requests.

## 5.2 Contribution of this Paper

This work contributes to the fields of Edge Computing and Computer Vision in IoT through the following:

**Feasibility Analysis:** Empirically evaluates CV workloads, highlighting latencies, energy consumption, and inference trade-offs.
**Communication Strategies:** Assesses LoRaWAN scenarios while demonstrating the satellite IoT concept again with presentable metrics as a complementary communication mean that can support the distributed and remote CV deployments.
**Integration Pathways:** Identifies how the emerging technologies such as MEC in 6G, multimodal AI and generative AI can extend the scalability and importance of edge-based CV systems.
**Socioeconomic Relevance:** Citing the potential impact of edge-enabled CV for precision agriculture, smart cities, environmental monitoring or industrial automation based on the resource constrained implementations and remote environments.
**Future Directions:** The covered interdisciplinary sources direct the research toward imaging and telecommunication aspects involving LEO, HAP, and LAP satellites to enable multimodal and generative AI at the edge. These developments will broaden the impact of CV in IoT.

## 6. Conclusions

The empirical findings of this study confirm the technical feasibility of Computer Vision (CV) within the Internet of Things (IoT) ecosystems. Evaluations are focused on three latency pillars: (1) the image acquisition and NN load time, (2) the DL inference duration and (3) the data parse and relay. Based on the measurements in Tables 6 and 7 the system I/O loading times span between 7 to 61 seconds, while the inference duration spanned between 30 and 138 seconds among the two platforms. Table 3 further demonstrates the uplink airtime ranged between 60 ms to 1 & ½ seconds attributed from SF and payload sizes. The experimentation use case for precision agriculture achieved encouraging results on important CV metrics: 0.76 on *precision*, 0.65 on *recall* and 0.70 on *f1*-score by using a modest-sized dataset. The robustness of the detection results is further enhanced by applying a confidence threshold, particularly under the dynamic occluded environments in an apple orchard. Computation availability under EC regimes is optimized using attributes such as the *energy-per-inference ratio*. This metric helps developers balance resource utilization with sustainability goals. On the communication spectrum, LoRaWAN demonstrates reliability in uplinks over extended distances, proving its utility in remote sensing applications. The findings support the protocol use alongside EC for scalable, energy-efficient deployments in various domains. A forward-looking discussion on Satellite IoT (SatIoT) emphasizes the relevance in the battery operated and geographically dispersed deployments. The broader implications of this study validate the real-world system behaviour, demonstrating the suitability for remote sensing applications where near real time response is sufficient. Further standardization as being addressed in **2.1.2**, play a pivotal role to ensure the interoperability across edge-IoT ecosystems. Efforts by ETSI (MEC specifications), IEEE (particularly for LPWAN and IEEE 802.15.4) and 3GPP (for NB-IoT and upcoming 6G) provide different frameworks for streamlining those protocols, interfaces, and data formats. Multimodal and generative AI at the edge highlights an emerging paradigm where intelligent systems are deployed locally to offer improved responsiveness. As advancements in hardware will continue to evolve with hybrid cloud models, the role of EC will support digital transformation, sustainability, and socioeconomic impact, positioning it as a cornerstone in future of IoT and AI-powered systems.